\documentclass[11pt,english]{article}

\usepackage[T1]{fontenc}
\usepackage[utf8]{inputenc}
\usepackage{lmodern}
\usepackage{amsmath}
\usepackage{amssymb}
\usepackage{cancel}
\usepackage[english]{babel}
\usepackage{physics}
\usepackage[toc,page]{appendix}
\usepackage{comment}
\usepackage{cite}

\usepackage{geometry}
\usepackage{caption}
\usepackage{subcaption}
\usepackage{graphicx}
\usepackage{physics}
\usepackage[export]{adjustbox}
\usepackage{caption}

\usepackage{epstopdf} 
\usepackage{comment}
\usepackage{tabularx} 
\usepackage{booktabs} 
\usepackage{psfrag} 

\usepackage{eurosym} 

\usepackage{xcolor} 
\usepackage{empheq} 

\definecolor{linkcolor}{rgb}{0,0,0.6} 
\usepackage[ 
    pdftex,
    colorlinks=true,
    pdfstartview=FitV,
    linkcolor=linkcolor,
    citecolor=linkcolor,
    urlcolor=linkcolor,
    hyperindex=true,
    hyperfigures=false,
    backref=page,            
]{hyperref}
\renewcommand*{\backref}[1]{}
\renewcommand*{\backrefalt}[4]{%
  \ifcase #1 %
    (Not cited.)%
  \or
    (cit. on p.~#2)%
  \else
    (cit. on pp.~#2)%
  \fi
}

\usepackage{fancyhdr} 
\usepackage{pgfplots} 
\usepgfplotslibrary{groupplots}

\newcommand{\go}[1]{\mathcal{O}(#1)}

\newcommand\cA{\mathcal{A}}
\newcommand\cB{\mathcal{B}}
\newcommand\cC{{\mathcal{C}}}

\newcommand\nth{{p}}

\newcommand\Ftwo{{F}}

\newcommand\cCo{{C}}
\newcommand\cDis{D}
\newcommand\XH{X_{\rm H}}
\newcommand\Xb{\bar{X}}

\numberwithin{equation}{section}
\numberwithin{figure}{section}
\numberwithin{table}{section}
\newcolumntype{Y}{>{\centering\arraybackslash}X}
\pgfplotsset{compat=1.6}
\pgfplotsset{/pgf/number format/use comma}

\author{Christos Charmousis$^{a,b}$, Simon Iteanu$^{a}$, David Langlois$^{c}$ and Karim Noui$^{a}$ \\
\small $^{a}${\it{Universit\'e Paris-Saclay, CNRS/IN2P3, IJCLab, 91405 Orsay, France}}\\
\small $^{b}${\it{Theoretical Physics Department, CERN, 1211 Geneva 23, Switzerland}}\\
\small $^{c}${\it{Université Paris Cit\'e, CNRS, Astroparticule et Cosmologie, F-75013 Paris, France}}\\
\tiny CERN-TH-2025-058
}

\date\today
\title{Axial perturbations of black holes with primary scalar hair}

\begin{document}

\maketitle
\begin{center}
{\bf Abstract}
\end{center}
We study axial perturbations of static black holes with primary hair in a family of degenerate higher-order scalar-tensor (DHOST) theories. These solutions  possess a scalar charge, fully independent of the mass,  leading to a continuous one-parameter deformation of the standard Schwarzschild black hole. Starting from these solutions, we also construct new  black holes, solutions of  other DHOST theories, obtained via disformal transformations of the metric. In particular, we investigate two specific types of disformal transformations: the first leading to a theory where gravitational waves propagate at the speed of light, the second to a Horndeski theory, where the equations of motion remain second order. 
 The dynamics of axial perturbations can be formally related to the general relativistic equations of motion of axial perturbations in an effective metric. The causal structure of the effective metric differs from that of the background metric,   leading to distinct gravitational and luminous horizons. Using a WKB approximation, we compute the quasi-normal modes for the Schrödinger-like equation associated with the effective metric outside the gravitational horizon.

\section{Introduction}

In recent years there has been an impressive amount of observational data on compact objects, namely neutron stars and black holes.
Indeed, it has almost become common practice to observe gravitational waves (GW)
emitted from mergers of black hole binaries (see for example \cite{LIGOScientific:2016aoc,LIGOScientific:2017vwq,LIGOScientific:2020zkf,LIGOScientific:2021qlt}). We also have images of supermassive black holes generated
from networks of radio-telescopes such as the Event Horizon Telescope (EHT) \cite{EventHorizonTelescope:2020qrl} and X-ray observations, like the NICER mission (Neutron star Interior Composition ExploreR), gathering information on the equation of state (EoS) of neutron stars by observing their
thermal hotspots \cite{Raaijmakers:2019dks}. 
The motion of stars in the vicinity of the supermassive black hole (BH) at the center of our Galaxy has been tracked by   the GRAVITY observatory since 2016, the precision ever improving with time~\cite{GRAVITY:2018ofz}. So far, observational results are quite compatible with predictions
emanating from General Relativity (GR)\footnote{Certain questions
do arise, regarding for instance the nature of the secondary object in 
GW190814 \cite{LIGOScientific:2020zkf}: its mass of $2.59_{-0.09}^{+0.08}~M_\odot$ places it in the current observational mass gap predicted
by GR, in between neutron stars and astrophysical black holes.}. Future observational data, including the next generation of GW detectors such as LISA \cite{Barausse:2020rsu} and the Einstein Telescope \cite{Abac:2025saz},
will permit to go further, by probing the validity of no-hair theorems,  
testing GR more drastically and putting tighter  constraints on alternative theories of gravity.

Among theories of modified gravity, DHOST theories \cite{Langlois:2015cwa,Langlois:2015skt,BenAchour:2016fzp} (see e.g. \cite{Langlois:2018dxi,Kobayashi:2019hrl} for reviews) encompass  most known scalar-tensor theories admitting a single scalar degree of freedom, in addition to the metric. They are therefore relatively simple, yet not trivial, modifications of GR and also include  limits of many  modifications of gravity which have a smooth GR limit, for example bigravity, massive gravity, braneworld theories, EFTs originating from string theories and so forth (see e.g.~\cite{Berti:2015itd}). DHOST theories therefore constitute a valuable bottom-up approach when considering geometric modifications of GR. 
 
Finding explicit solutions beyond GR, in particular in DHOST theories, is a technically involved task (for a recent review see for example \cite{Babichev:2023psy, Lecoeur:2024kwe}) and often relies on the scalar and the geometry admitting some specific symmetries. Exact solutions of black holes are particularly interesting as they often provide in-depth insights about the underlying theory, something which is more difficult to extract from numerical simulations. For example,  they help address questions such as understanding the interior structure beyond the event horizon or the nature of the central singularity, evaluating explicitly geodesics and analysing their properties and their mathematical integrability, or computing linear perturbations and identifying their differences with GR. As such, several explicit BH solutions were found in the last decade or so, especially in the case of higher order scalar tensor theories ranging from Horndeski to the most general case of DHOST theories.

Early on, stealth solutions, i.e solutions with a GR metric and  a non trivial scalar field profile, were found for theories with a global shift symmetry for the scalar field. In certain cases 
these solutions provide a self-tuning of the  cosmological constant~\cite{Babichev:2013cya, Kobayashi:2014eva, Charmousis:2015aya, Babichev:2016kdt}, although they are strongly constrained when taking into account the simultaneous measurement \cite{LIGOScientific:2016aoc, LIGOScientific:2017vwq} of the speed of gravitons and light (for an analysis see \cite{Babichev:2017lmw}). These early stealth solutions were extended by finding the first rotating solution, stealth Kerr \cite{Charmousis:2019vnf}, which was later promoted to a genuinely different black hole  geometry other than Kerr, the disformed Kerr metric \cite{Anson:2020trg,BenAchour:2020fgy}. More recently, using local conformal invariance for the scalar, it was realised that one could construct black holes originating from higher dimensional Lovelock theory{\footnote{For the full analysis in this direction see \cite{Colleaux:2019ckh}.}} with especially interesting properties \cite{Fernandes:2021dsb, Charmousis:2021npl, Fernandes:2023vux}. Finally, certain solutions were constructed in \cite{Babichev:2023dhs}, which have no explicit symmetry; rather they have some remnant symmetry originating from higher dimensions.

 Recently, it was shown in \cite{Bakopoulos:2023fmv} that a family of shift symmetric DHOST theories  allow quite simple, well-defined black holes with primary hair (see furthermore the extensions found in \cite{ Baake:2023zsq, Bakopoulos:2023sdm}). In addition to the usual ADM mass parameter, there is an  independent integration constant associated with the scalar field, which gives rise, in certain cases, to a primary scalar Noether charge \cite{Bakopoulos:2023sdm}. This charge is linked to the shift symmetry of the theory in question, allowing for a linear time dependence for the scalar field, as initially introduced in \cite{Babichev:2013cya}. The scalar field therefore, unlike the metric, is not static, but crucially the energy momentum tensor associated with the scalar field is. This simple trick  evades the typical staticity hypothesis of no hair theorems \cite{Hui:2012qt, Babichev:2016rlq} without tempering mathematical robustness.

These black holes have several interesting properties. First of all, when the scalar charge is turned off, one recovers a Schwarzschild BH, thus indicating a regular GR limit  for all solutions (unlike for example BCL \cite{Babichev:2017guv},  or without the need to go to another branch of solutions such as in \cite{Doneva:2017bvd,Silva:2017uqg,Antoniou:2017acq}).
 Secondly, the scalar field is regular at all future directed horizons  for all values of the scalar charge. This is a crucial property as the charge is not fixed by requiring regularity at the horizon, in which case it would be downgraded to secondary hair (i.e. it would depend on the particular BH at hand,  in particular on its mass). Thirdly, for a particular value of the scalar charge, which depends on the ADM mass, the solution turns out to be regular, i.e. without any central spacetime singularity. The solutions can therefore give regular black holes or solitons (if there is no horizon) without requiring any fine tuning of the theory. 
 
Starting from any BH solution in a DHOST theory, one can easily construct new solutions via conformal-disformal transformations of the metric, which can be seen as a field redefinition of the metric~\cite{BenAchour:2016cay}. The disformed metric then provides a new BH solution associated with a different DHOST theory, provided ordinary matter and light is minimally coupled to the disformed metric. In the present work, we construct explicitly such  new solutions from the BHs found in \cite{Bakopoulos:2023fmv,Baake:2023zsq,Bakopoulos:2023sdm}, emphasizing two particular cases: when the new theory is such that the axial modes propagate with the speed of light and when the new theory belongs to the Horndeski subclass of DHOST theories.

Given these well-defined black hole solutions, a pressing question is whether they are stable, at least with respect to linear perturbations. Moreover, for stable black holes, it is instructive to compute their oscillating modes, since future observations of binary BH  mergers via GWs should be able to detect the oscillations of the newly created BH  in the  post-merger ringdown phase. The main contribution to these oscillations can be decomposed into discrete modes known as quasi-normal modes (QNMs) – instead of normal modes because these modes decay as they are radiated away. The measurement of the QNM frequencies and their associated decay rates  could provide a very powerful test of general relativity (GR) in strong gravity, as alternative theories of gravity predict different  QNMs with respect to GR. 

Several works  have investigated linear perturbations of nonrotating black holes in  DHOST theories or particular  subfamilies~\cite{Kobayashi:2012kh,Cisterna:2015uya,Takahashi:2016dnv,Takahashi:2019oxz,deRham:2019gha,Charmousis:2019fre,Khoury:2020aya,Tomikawa:2021pca,Langlois:2021xzq,Langlois:2021aji,Takahashi:2021bml,Chatzifotis:2021pak,Minamitsuji:2022mlv,Langlois:2022eta,Minamitsuji:2022vbi,Langlois:2022ulw,Noui:2023ksf,Roussille:2023sdr,Antoniou:2024hlf,Antoniou:2024gdf}. An EFT treatment of these perturbations has also been developed~\cite{Franciolini:2018uyq,Hui:2021cpm,Mukohyama:2022enj,Mukohyama:2023xyf,Mukohyama:2025owu}.
For spherically symmetric black holes, the QNMs can be divided into two classes of modes: polar (even-parity) modes and axial (odd-parity) modes. 
In the present work, we will focus our attention  on the axial modes of the black holes with primary hair discussed above. In scalar-tensor theories, the axial modes are simpler than the polar modes as they do not involve a perturbation of the scalar field. 
As shown in \cite{Langlois:2022ulw} (see also \cite{Tomikawa:2021pca}), there is a nice correspondence between the dynamics of axial modes in quadratic DHOST theories and that of axial modes in GR but in a different metric, which we call the effective metric. 

Interestingly, this effective metric can be understood as a specific conformal-disformal transformation of the background metric into what  we refer to as the Einstein frame\footnote{The Einstein frame here corresponds to the theory where the dynamics of axial modes is identical to that in GR.}. If the background metric corresponds to a black hole, the effective metric may also describe a black hole with a potentially modified horizon. Remarkably, the two geometries can, in some cases, be of different types: for instance, an effective black hole can coexist with a solitonic background solution. We apply the usual procedure to derive a Schrödinger-like equations governing the axial perturbations propagating in the effective metric and, using a WKB approximation, we compute   numerically the lowest quasinormal modes for several examples of effective metric. This allows us to quantify how the corresponding QNM frequencies  vary with the scalar charge associated with the primary hair.

The outline of the paper is the following. In the next section, we introduce the specific DHOST theories that we consider in this work and  present several explicit solutions in these theories. In section \ref{Section_disformal}, we show how to construct  new black hole solutions by using disformal transformations of the metric. Section \ref{Section_axial} is devoted to axial modes and the effective metric in which the pertubations propagate. We then compute numerically, in Section \ref{Section_QNMs}, the lowest QNMs for several BH solutions. The final section summarises our results and discusses certain open issues.  We have also added some appendices where more detailed calculations are provided.

\section{Black holes with primary hair}
\label{BH_primary_hair}

In this section, after briefly presenting  DHOST theories, we review some exact static black hole solutions  with primary hair that have recently been discovered in \cite{Bakopoulos:2023fmv} and, soon after,  extended in \cite{Baake:2023zsq} and \cite{Bakopoulos:2023sdm}. 

\subsection{Theory and homogeneous solutions with primary hair}
We consider  DHOST theories (up to quadratic terms in the second derivatives of the scalar field) that are described by the action \cite{Langlois:2015cwa}
\begin{align}
\label{DHOSTaction}
S\left[g_{\mu \nu}, \phi\right]= \int \mathrm{d}^4 x \sqrt{-g} \left( P(X, \phi)+ Q(X, \phi) \square \phi+\Ftwo(X, \phi) R +  \sum_{i=1}^5 A_i(X, \phi) L_i^{(2)} \right)   
\end{align}
where the kinetic density
\begin{equation}
    X\equiv  -\frac{1}{2} \partial_\mu \phi \, \partial^\mu \phi\,.
\end{equation}
Many of our equations are also written with the alternative convention $X_{\rm other}\equiv \partial_{\mu}\phi\, \partial^\mu \phi$ in  Appendix \ref{App:X_other}. 
In the above action, $P$, $Q$, $\Ftwo$ and $A_i$ are functions of $\phi$ and $X$ while the expressions of the elementary quadratic Lagrangians can be found  in \cite{Langlois:2015cwa}. The functions $\Ftwo$ and $A_i$ must satisfy three so-called degeneracy conditions so that the theory contains a single scalar degree of freedom. 

The theories that will be discussed in this paper all belong to a sub-class of
DHOST theories, known as Beyond Horndeski theories~\cite{Gleyzes:2014dya,Gleyzes:2014qga}, characterised by the conditions
\begin{eqnarray}
\label{Ai_GLPV}
A_2=-A_1   \,, \qquad A_4=-A_3= \frac{A_1+\Ftwo_{,X}}{X} \,, \qquad A_5=0\,,
\end{eqnarray}
which ensure that the degeneracy conditions are verified. This sub-class itself contains the Horndeski theories~\cite{Horndeski:1974wa}, which satisfy, in the DHOST notation, the more restrictive conditions
\begin{eqnarray}
\label{Ai_Horndeski}
A_2=-A_1= \Ftwo_{,X}  \,, \qquad A_3=A_4=A_5=0\,.
\end{eqnarray}
From now on, we restrict our attention to Lagrangians in which $Q=0$ and the remaining independent functions $P$, $\Ftwo$ and $A_1$ depend only on $X$, which implies in particular that  the theory has shift ($\phi \rightarrow \phi + c$ where $c$ is a constant) and parity global symmetry ($\phi \rightarrow -\phi$) for the scalar field $\phi$.

\medskip
Static and  spherically symmetric solutions in the above theories can be described by a metric of the form
\begin{align}
\label{background_metric}
\mathrm{d} s^2 =-\cA(r) \mathrm{d} t^2+\frac{\mathrm{d} r^2}{\cB(r)} +r^2 \mathrm{d} \Omega^2\,, \qquad \mathrm{~d} \Omega^2=\mathrm{d} \theta^2+\sin ^2 \theta \mathrm{d} \varphi^2\,,
\end{align}
while, in accordance with the global shift symmetry of the theory, the scalar field can display a linear time dependence,
\begin{align}
\label{scalaransatz}
    \phi(t,r)=q t+\psi(r)\,,
\end{align}
where $q$ is an integration constant. 
It has been shown \cite{Bakopoulos:2023fmv} that the field equations reduce to three simple independent equations,\begin{equation}
\label{easy}
 \frac{\cA}{\cB}=\frac{\gamma^2}{Z^2},
\end{equation}
\begin{equation}
\label{easy2}
    r^2(P Z)_{X}+2(\Ftwo Z)_X \left(1-\frac{q^2 \gamma^2}{2 Z^2 X}\right)=0,
\end{equation}
\begin{equation}
\label{easy3}
    2\gamma^2 \left(\cA r-\frac{q^2 r}{2 X}\right)'=-r^2 P Z-2\Ftwo Z\left(1-\frac{q^2 \gamma^2}{2 Z^2 X}\right)+\frac{q^2\gamma^2X' r}{ZX^2}\left( 2X\Ftwo_{X}-\Ftwo\right),
\end{equation}
where a subscript $X$ denotes a derivative  with respect to $X$ while a prime denotes a derivative with respect to the radial coordinate $r$. The three equations include an integration constant $\gamma$ which we can fix to unity without loss of generality and $Z$ is tailored to the theory via{\footnote{In beyond Horndeski notation, we have $Z=-G_4+2 X G_{4X}+4 X^2 F_4$.}}
\begin{equation}
\label{Z}
    Z=-\Ftwo-2 X A_1 \, .
\end{equation}
Hence \eqref{easy} tells us that homogeneous black holes, i.e. such that  $\cA=\cB$, are only possible{\footnote{We choose the negative sign as the definition \eqref{Z} dictates that $Z$ corresponds to $-F$ in the absence of $A_1$. Hence the GR limit corresponds to $F=1$.}} for $Z=-1$  while any other non homogeneous solution will be parametrised by a non trivial $Z$ whose expression depends on the given theory.

Let us first concentrate on homogeneous solutions with $Z=-1$. The second  equation above, Eq.~\eqref{easy2}, does not involve the metric coefficients, hence it is an algebraic equation for $X$ and effectively gives the scalar field. Explicit solutions have been obtained for a family of  DHOST theories \eqref{DHOSTaction} whose associated functions\footnote{This family belongs to Beyond Horndeski theories. If one uses the so-called "beyond Horndeski notations" for the action, as it was done in the original papers \cite{Bakopoulos:2023fmv,Baake:2023zsq,Bakopoulos:2023sdm}, we have
    $$G_2(X)=-\frac{2 \alpha}{ \lambda^2} X^\nth,\; \quad
    G_4(X)=1- \alpha X^\nth,\; \quad F_4(X)=\frac{\alpha}{4}(2\nth-1) X^{\nth-2} \, .$$.} take the form
\begin{align}
\label{P_n}
    P(X)&=-\frac{2 \alpha}{ \lambda^2} X^\nth \, , \quad
    \Ftwo(X)=1- 2X A_1(X)\, , \quad
    A_1(X) =\frac{\alpha}{2} X^{\nth-1}\,, \quad 
A_3(X)=\frac{\alpha}{2}(2\nth-1) X^{\nth-2}\,,
\qquad
\end{align}
while $A_2=-A_1$, $A_4=-A_3$ and $A_5=0$, thus  belonging to beyond Horndeski theories according to \eqref{Ai_Horndeski}.
These theories are characterised by three parameters: $\nth$   takes positive integer or half-integer values henceforth, $\alpha$  is a dimensionless coupling constant and $\lambda$  is a constant with the dimension of a length. 
It is immediate to see that indeed $Z=-1$ for this class of theories. 

For any $\nth$ and $\alpha$, the second field equation \eqref{easy2} leads to
the following solution for $X$ from which we can easily extract $\psi'(r)$,
\begin{equation}
X=\frac{q^2 /2}{1+(r / \lambda)^2}, \qquad \psi^{\prime}(r)^2=\frac{q^2}{\cA(r)^2}\left[1-\frac{\cA(r)}{1+(r / \lambda)^2}\right] \,. \label{scalar}
\end{equation}
The last equation \eqref{easy3} is a simple ordinary differential equation for the metric component $\cA$.
It can be solved for the family of theories \eqref{P_n} considered here and, using \eqref{scalar}, it gives
\begin{equation}
    \cA(r)=1-\frac{2\mu}{r}-\xi_\nth\frac{2\lambda}{r} \,\Xi_\nth(r/\lambda)\,,\qquad \Xi_\nth(x)\equiv\int_0^x \mathrm{d}u \; \frac{u^2}{\left(1+u^2\right)^\nth}\,,
    \label{solAq0}
\end{equation}
where $\mu$ is an integration constant (with dimension of length) and  $\xi_\nth$ is a dimensionless parameter, 
\begin{equation}
\label{xi}
    \xi_\nth\equiv \alpha ({2\nth-1}) \left({q^2}/{2}\right)^{\nth} \,,
\end{equation}
which can be interpreted as the ``strength'' of the scalar hair.
The integral $\Xi_\nth(r)$ corresponds to an hypergeometric function \cite{Baake:2023zsq,Bakopoulos:2023sdm},
\begin{equation}
    \Xi_p(x)=\frac{x^3}{3}\, {}_2F_1(3/2,p;5/2;-x^2)\,,
\end{equation}
which can be written in an explicit form for  special values of $p$, as will be illustrated below. When $x\to +\infty$, $\Xi_p$ behaves as 
\begin{equation}
    \Xi_p(x)= \frac{\sqrt{\pi}\,\Gamma(p-\frac32)}{4\, \Gamma(p)}+{\cal O}(\frac{1}{x^3})+x^{-2p}\left(\frac{x^3}{3-2p}+\frac{p}{2p-1}x-\frac{p(1+p)}{2(1+2p)x}+{\cal O}(\frac{1}{x^3})\right)\,,
\end{equation}
for $p\neq -1/2, 1/2, 3/2$ (the special case $p=1/2$ is discussed below). It is  convenient to combine the constant term of $\Xi_p(r/\lambda)$, in this asymptotic regime, with $\mu$ and define the constant
\begin{equation}
    M\equiv \mu+\frac{\sqrt{\pi}\,\Gamma(\nth-\frac32)}{4\, \Gamma(\nth)}\xi_\nth\qquad (\nth\neq -1/2, 1/2, 3/2)\,,
\end{equation}
which can be interpreted as the usual ADM mass of the solution (note that $M$ has the dimension of length, since we work implicitly in units where $G=1$ and $c=1$).

In summary, setting :
\begin{equation}
    \check{\Xi}_\nth=\frac{\mu -M}{\xi_\nth}+\Xi_\nth,
\end{equation}
we have  obtained a solution with primary hair, of the form
\begin{equation}
    \cA(r)=1-\frac{2M}{r}-\xi_\nth\frac{2\lambda}{r} \, \check{\Xi}_\nth(r/\lambda)\,,
    \label{solAq}
\end{equation}
parametrised by the mass $M$ and the constant $q$ (hidden in $\xi_\nth$), which quantifies the deviation of the solution from the standard Schwarzschild  black hole corresponding to the value $q=0$. It is worth stressing that the solution is not necessarily a black hole. Indeed,  if the function $\cA$ does not vanish, the solution describes a naked singularity, or a soliton in the particular cases where $\cA$ is regular at the origin. Since $\Xi_\nth(r/\lambda)\simeq (r/\lambda)^3/3$ when $r\to 0$, it is clear from \eqref{solAq0} that the solution is regular for $\mu=0$, i.e. when the mass $M$ takes the particular value
\begin{equation}
\label{M_reg}
    M^{\rm reg}_\nth\equiv \frac{\sqrt{\pi}\,\Gamma(\nth-\frac32)}{4\, \Gamma(\nth)}\xi_\nth\qquad (\nth\neq -1/2, 1/2, 3/2)\,.
\end{equation}
If $\cA$ vanishes for some finite radius, then the solution with the mass $M^{\rm reg}_\nth$ yields a regular black hole, i.e. devoid of singularity behind the horizon.

\medskip
Before considering explicit examples in the following sub-section, let us discuss some properties of the scalar hair. It was noted in \cite{Bakopoulos:2023sdm} that the parameter $q$ is closely related to a Noether charge that comes with the Noether current $J^\mu$  associated with the shift symmetry of the scalar field.
 Indeed, the only non-trivial (on-shell) component of the current is $J^t$ that can be written as follows (up to an overall irrelevant factor), 
\begin{eqnarray}
    J^t = \frac{q^2(4 (p-1) p+3)  +4 X p (1-2p) }{\lambda ^2 q^3} X A_1(X)\, .
\end{eqnarray}
Hence, following \cite{Bakopoulos:2023sdm} one can immediately compute the associated charge $Q$ which is given (up to an overall irrelevant factor) by the sum of two integrals

\begin{eqnarray}
\label{charge}
    Q \propto   \; \frac{\xi_p}{\lambda^2 q} \left[
    \frac{4 (p-1) p+3}{2 (2\nth -1)}   \, \Xi_p^\infty  \,  - \, \nth \, \Xi_{p+1}^\infty  \right]\, ,
    \qquad \Xi_p^\infty = \lim_{r \to +\infty} \Xi_p(r) = \frac{\sqrt{\pi}\,\Gamma(p-\frac32)}{4\, \Gamma(p)}\, . 
\end{eqnarray}
 The above
limit  $\Xi_p^\infty$ converges, which means that the charge is well-defined,  only for $\nth>3/2$. In that case, the charge $Q$ is
intimately linked to the hair parameter $q$.

\subsection{Examples of homogeneous black holes}
\label{sec: examples}

In the following, we will set $\lambda=1$ in order to shorten our notation. It is easy to reintroduce the length parameter $\lambda$ in all expressions by applying the substitutions $r\to r/\lambda$ and $M\to M/\lambda$.
We now focus on some explicit examples for values of $\nth$ that present most interest.

\subsubsection*{Case $\nth=1/2$}
This special value of $\nth$ corresponds to a pure Horndeski theory, i.e. satisfying \eqref{Ai_Horndeski}, with
\begin{equation}
\label{1/2}
P(X)=-{2\alpha} \sqrt{X}, \quad \Ftwo(X)=1-\alpha\sqrt{X} , \quad A_1(X)=\frac{\alpha}{2\sqrt{X}}\,,
\end{equation}
In this case, the spacetime metric is identical to the GR Schwarzschild metric, since $\xi_{1/2}=0$ and \eqref{solAq0} reduces to
\begin{equation}
\label{stealth_metric}
\cA(r)=1-\frac{2 M}{r}\,,
\end{equation}
while the scalar field is given by $\phi=qt+\psi(r)$ and \eqref{scalar}.
This is the only case where the scalar parameter $q$ does not appear in the spacetime metric and the metric is therefore stealth Schwarzschild \cite{Bakopoulos:2023fmv}. This is a complementary case to the stealth Schwarzschild solution first found in \cite{Babichev:2013cya} with the notable difference that $X$ there was constant. Another important difference is that, even when $M=0$, we find a flat spacetime metric with a non trivial and regular scalar field, in other words a flat soliton solution given by
\begin{equation}
\label{soliton}
 \phi=q t+  \frac{q}{\sqrt{1+r^2}}\, .
\end{equation}
This stems from the fact that $X$ is always regular and independent of the mass parameter. 

\subsubsection*{Case $\nth=1$}
The case $\nth=1$ is  characterised by the functions
\begin{align}
    P(X)=-{2 \alpha} X\,, \quad
    \Ftwo(X)=1- \alpha X\,\quad
    A_1(X)=\frac{\alpha}{2}\,,\quad
A_3(X)=\frac{\alpha}{2 X}\,,
\end{align}
and therefore the theory belongs to the beyond Horndeski class \eqref{Ai_GLPV}. What makes this case special is that there is canonical kinetic term in the action. 
The metric component takes the form~\cite{Bakopoulos:2023fmv}
\begin{equation}
\label{n1}
 \cA(r)=1 -\frac{2 M}{r}-2\xi_1\left(1+\frac{\pi / 2-\arctan r}{r }\right)\,,
\end{equation}
while the scalar field expression follows from  \eqref{scalar} with $\cA$ now given above.

In this particular case, we see that $\cA(r)=1-2\xi_1+\go{1/r}$ at spatial infinity, which means that
the solution is only {\it locally} asymptotically flat. Indeed, in the limit where $r \rightarrow \infty$, the spacetime geometry is similar to that of a gravitational monopole with a solid angle deficit or excess, depending on the sign of $\xi_1$. This can result in strong lensing effects for objects residing behind the black hole along the line of sight. 
Moreover to avoid the pathological cases where ${\cA}$ is everywhere negative, we
should require $\xi_1$ to satisfy the condition $\xi_1<1/2$.

Finally, according to \eqref{M_reg}, the solution becomes  regular, i.e. is free from the central singularity, for the mass
\begin{equation}
\label{regular1}
M^{\rm reg}_1=-\frac{\pi \xi_1 }{2}\,,   
\end{equation}
which is possible only if $\xi_1<0$. And this regular solution is always a soliton, not a black hole.

\subsubsection*{Case $\nth=2$}
For $\nth=2$, the theory  is defined by the  functions 
\begin{equation}
\label{n2}
P(X)=-{2 \alpha}X^2, \quad \Ftwo(X)=1-\alpha X^2\, ,\quad A_1(X)=\frac{\alpha}{2} X\,,\quad
A_3(X)=\frac32\alpha\,,
\end{equation}
yielding the metric coefficient 
\begin{equation}
\label{A_n2}
\cA(r)=1-\frac{2 M}{r}+\xi_2\left(\frac{\pi / 2-\arctan r}{r }+\frac{1}{1+r^2}\right) \, .
\end{equation}
In contrast to the previous case, the solution is asymptotically flat.

The solution becomes regular when the mass $M$ takes the special value
\begin{equation}
\label{Mreg2}
    M^{\rm reg}_2=\frac{\pi \xi_2 }{4}\,,
\end{equation}
which requires $\xi_2>0$. Interestingly, depending on the value of $\xi_2$, the regular solution can be either a black hole, if $\xi_2> \xi_{2,{\rm sol}}\simeq 2.816$, or a soliton for smaller values of $\xi_2$  (see Appendix \ref{App:phase_diagram} for the computation of $\xi_{2,{\rm sol}}$). 

\subsubsection*{Case $\nth=5/2$}

Another interesting case corresponds to $\nth=5/2$ where the  functions of the theory are given by
\begin{equation}
\label{n52}
P(X)=-{2 \alpha} X^{5/2}, \quad \Ftwo(X)=1-\alpha X^{5/2}, \quad A_1(X)=\frac{\alpha}{2} X^{3/2}\, ,\quad
A_3(X)=2\alpha\sqrt{X}\,.
\end{equation}
The homogeneous black hole  is now described by the metric coefficient~\cite{Bakopoulos:2023sdm,Lecoeur:2024kwe}
\begin{equation}
\label{A_n52}
\cA(r)=1-\frac{2 M}{r}+\frac{2\xi_{_{5/2}} }{3r}\left(1-\frac{r^3}{(1+r^2)^{3/2}}\right) \,,
\end{equation}
which becomes regular  for the special value of the  mass given by
\begin{align}
    M^{\rm reg}_{5/2}\equiv\frac{ \xi_{_{5/2}}}{3} \,.
\end{align} 
Again, the regular solution describes a  soliton for small values of $\xi_{_{5/2}}$, namely $\xi_{_{5/2}}<\xi_{_{5/2},{\rm sol}}=9\sqrt{3}/4$, and a black hole for larger values (see Appendix \ref{App:phase_diagram} for details).

\subsubsection*{``Adding" solutions}

It is interesting to note that in the space of theories \eqref{P_n} the primary hair sources can be superposed.  Indeed, we can ``add'' two such theories by summing the functions $P$ and $A_i$ without changing the first two equations \eqref{easy} and \eqref{easy2}. As a consequence, the sum of two theories still admits a similar solution with a new metric that can be interpreted as the ``sum'' of the two.

Furthermore, a square root term associated to the theory $\nth=1/2$, Eq.  \eqref{1/2}, can be included in any theory without altering the black hole solution itself. For example the theory described by
\begin{eqnarray}
\label{yo}
&&P(X)=-2\alpha \sqrt{X}-2 \alpha X^{5/2},\qquad\Ftwo(X)=1-2XA_1(X) \; , \nonumber \\
&&A_1(X)=\frac{\alpha}{2\sqrt{X}}+\frac{\alpha}{2} X^{3/2},\qquad
A_3(X)=2\alpha\sqrt{X} \, ,
\end{eqnarray}
admits the very same solution \eqref{A_n52}.

\section{Disformal transformations and non homogeneous black holes}
\label{Section_disformal}
In this section, we construct new non homogeneous black hole solutions making use of disformal transformations~\cite{Bekenstein:1992pj}. Such a technique has shown to be very useful to find new solutions in DHOST theories \cite{BenAchour:2020wiw}.

\subsection{Generic conformal-disformal  transformations}

As shown in \cite{BenAchour:2016cay}, a conformal-disformal  transformation  of the  metric,
\begin{align}
\label{disformal_transf}
    \tilde g_{\mu\nu}= \cCo(X,\phi) \, g_{\mu\nu} + \cDis(X,\phi) \, \partial_\mu \phi\, \partial_\nu \phi\,,
\end{align}
where $C$ and $D$ are functions of $X$ and $\phi$,
induces an internal map in the space of DHOST theories. Indeed, any DHOST theory governed by an action $\tilde S[\tilde{g}_{\mu\nu},\phi]$ is related to another DHOST theory governed by the action
\begin{eqnarray}
    S[g_{\alpha\beta},\phi] \equiv \tilde{S}[\tilde{g}_{\mu\nu}(g_{\alpha\beta},\phi),\phi]\,.
\end{eqnarray}
The explicit disformal transformations between quadratic DHOST theories can be found in the appendix D of \cite{Langlois:2020xbc} (correcting some typos in \cite{BenAchour:2016cay}). 

We stress that, while the seed and image theories are equivalent theories, provided the disformal transformation is invertible\footnote{The theories are inequivalent if the disformal transformation is non-invertible, which leads to mimetic-like theories, as discussed in \cite{Langlois:2018jdg} for instance.}, they become physically inequivalent  if one introduces matter and assumes it is  minimally coupled to their respective metric. In particular, in contrast with purely conformal transformations, disformal transformations (with a non trivial $D$ function)  lead to a new metric with a different light cone structure. 

While, in a generic quadratic DHOST theory (of type Ia), gravitational waves propagate with a speed $c_g$ that differs from that of light $c$, it is in general possible to find a disformally related  family of frames (or theories) where gravitational waves propagate at exactly the speed of light. All the theories in this family, which we will call the $c_g=c$ frame family, are obtained from the initial theory via disformal relations and  are related between themselves via purely conformal transformations, which automatically preserve the defining condition $c_g=c$. 

Similarly, any DHOST theory (again of type Ia) can be related, via disformal transformations, to a Horndeski frame family where all theories belong to the Horndeski class, with the interesting property that all the equations of motion are second order field equations.

\medskip

Let us now  compute explicitly the disformed metric $\tilde{g}_{\mu\nu}$ \eqref{disformal_transf} obtained from the spherically symmetric metric  \eqref{background_metric}. Using  the scalar field \eqref{scalar}, we  obtain  the new static and spherically symmetric metric
\begin{eqnarray}
\mathrm{d}{\tilde s}^2 &=& -(\cCo \cA - \cDis q^2)  \left( \mathrm{d}{t} - \cDis \frac{ q \psi'}{\cCo \cA - \cDis q^2} \mathrm{d}{r} \right)^2 + \cCo \left( \frac{1}{\cB} +  \frac{\cDis \cA\,  \psi'^2}{\cCo \cA - \cDis q^2}\right)  \mathrm{d}{r}^2 + 
\cCo\,  r^2  \mathrm{d}{\Omega}^2 \, .
\end{eqnarray}
We can easily diagonalise the above metric by introducing the new
time variable
\begin{equation}
\label{t_star}
    t_*=t-q\int\frac{\cDis\psi' }{\cCo \cA - \cDis q^2}\mathrm{d} r\,,
\end{equation}
which leads to
\begin{eqnarray}
d\tilde{s}^2 = -\tilde\cA(r) \, \mathrm{d}t_*^2+\frac{\mathrm{d} r^2}{\tilde\cB(r)} +\tilde\cC(r) \, \mathrm{d} \Omega^2 \,  ,
\end{eqnarray}
with
\begin{eqnarray}
 &&\tilde\cA= \cCo \left(\cA - q^2{\cDis}/{\cCo }\right) \, , \quad \frac{1}{\tilde \cB} =  \cCo \, \frac{\cA}{\cB} \left(   \frac{1 -2X{\cDis}/{\cCo }}{\cA - q^2{\cDis}/{\cCo }}\right) \, , \quad
\tilde\cC= \cCo r^2 \, .
\label{disformed_metric}
\end{eqnarray}
In terms of the new time coordinate \eqref{t_star}, the scalar field is expressed as  
\begin{equation}
    \phi =q t_*+\int\mathrm{d} r  \; \frac{\cA}{\cA -\frac{\cDis}{\cCo } q^2} \psi'\,.
\end{equation}
Inserting $\cA=\cB$ in \eqref{disformed_metric}, we see that a disformal transformation maps the original {\it homogeneous} black hole solution into a {\it non-homogeneous}  (i.e. $\tilde{\cA}\neq \tilde{\cB}$), corresponding to a non-constant function $\tilde{Z}=-\tilde{\Ftwo} - 2 \tilde{X} \tilde{A_1}$, as mentioned in \cite{Bakopoulos:2023sdm}. 

\medskip

By construction, the new  metric is a solution of a DHOST theory $\tilde{S}$ obtained from the original one $S$ by a disformal transformation. The relations between the ``coupling'' functions in these two actions can be found\footnote{Beware that $X$ in those references is defined as $X_{\rm other}\equiv \partial_\mu \, \phi \partial^\mu \phi$.} in \cite{BenAchour:2016cay} (see also the Appendix D in \cite{Langlois:2020xbc}). Here, we will just need the new functions $\tilde \Ftwo$ and $\tilde A_1$, given respectively by
\begin{eqnarray}
\label{disformed_F2}
\tilde \Ftwo & = & 
\frac{\Ftwo}{\cCo (1 -2X \cDis/\cCo)^{1/2}}
 \, ,\\
 \label{disformed_A1}
\tilde A_1 & = & \left( 1 -2X {\cDis}/{\cCo} \right)^{3/2} (A_1 - \frac{\cDis}{\cCo-2 X\cDis } \Ftwo) 
\,,
\end{eqnarray}
while the relation between $P$ and $\tilde{P}$ is the same as in \eqref{disformed_F2}.
It is also useful to recall the relation between $X$ and $\tilde{X}$,
\begin{align}
\label{disformed_X}
    \tilde{X}=\frac{X}{\cCo-2X\cDis}\,,
\end{align}
which can be inverted and substituted into \eqref{disformed_F2} and \eqref{disformed_A1} so that $\tilde \Ftwo$ and $\tilde A_1$ are expressed as functions of $\tilde{X}$. 

Finally, let us note that
 the transformations \eqref{disformed_F2} and \eqref{disformed_A1} are well defined only if $C$ and $D$ satisfy the condition 
\begin{eqnarray}
\label{CondDis}
    C(X)-2X D(X) \, > \, 0 \, , 
\end{eqnarray}
which in turn ensures that $\tilde{X}$ has the same sign as $X$. Furthermore, as can be seen from the expressions of the metric coefficients \eqref{disformed_metric}, this condition, together with $C>0$, also ensures that $g_{\mu\nu}$ and $\tilde{g}_{\mu\nu}$ keep the same global signature, even if  the signs of $\cA$ and $\cA - q^2{\cDis}/{\cCo }$ might differ (as will be sometimes the case in our examples). 

\subsection{Asymptotic properties of the disformed metrics}
\label{Someprop}

Before analysing in detail  the various cases, it is instructive to discuss in general the asymptoptic properties of the non-homogeneous metrics \eqref{disformed_metric} obtained by disformal transformations. Since a purely conformal transformation does not modify the causal structure of the metric, we fix here the conformal factor to $C=1$ for simplicity. 

Let us start by studying the behaviour of the disformed metric at the origin. Since $X=q^2/2+ \go{r^2}$ when $r \to 0$, $D$ tends to $D_0\equiv D(q^2/2)$, which must be finite, and  we thus find, using \eqref{solAq0}, the behaviour
\begin{eqnarray}
\label{cAatr0} 
    \tilde{\cA}  = 1-q^2 D_0-\frac{2\mu}{r}   + O(r^2) \,, \qquad
    \tilde{\cB}  =  1- \frac{2\mu}{1-q^2D_0} \frac{1}{r}+ O(r^2) \, . 
\end{eqnarray}
Note that the condition \eqref{CondDis} applied to $r=0$ leads to $q^2D_0<1$. Moreover, one sees that the regularity of the metric is conserved by the disformal transformation since both initial and disformed metric are regular when $\mu=0$. Indeed, in this case, $\tilde\cA$ is finite and
$\tilde \cB = 1$  at the origin\footnote{Indeed, the property $\tilde{\cB}(0)=1$ is crucial to make the space-time regular at the origin.}. However, the nature of the regular metric, soliton or regular black hole, is not necessarily preserved as we will see later in our explicit examples. 

Let us now examine the limit $r \to \infty$. If  the seed metric is asymptotically flat, which is always the case when $\nth\neq 1$ \eqref{n1}, we find
\begin{eqnarray}
\label{disformed_infinity}
    \tilde{\cA}  \to    1 -q^2 D_\infty\,,\qquad
    \tilde{\cB} \to 1 -q^2 D_\infty\qquad (r\to+\infty) \, ,
\end{eqnarray}
where $D_\infty = \lim\limits_{X \to 0} D(X)$. 
We therefore conclude that the disformed metric 
is globally asymptotically flat if $D_\infty =0$ and only locally asymptotically flat if $D_\infty$ is finite and non zero.
The particular case $\nth=1$ will be considered later.

\subsection{Solutions in the $c_g=c$  frame}
\label{sectioncgc}
As discussed earlier, it is possible to obtain disformed solutions in theories such that $c_g=c$. As shown in \cite{Langlois:2017dyl}, such a frame is characterised  by the condition $\tilde A_1=0$. Hence from the expression of $\tilde A_1$ given in \eqref{disformed_A1}, the disformal coefficient $D$  needed to reach this subfamily of theories must satisfy
\begin{align}
    \frac{\cDis}{\cCo}=\frac{A_1}{\Ftwo+2X A_1}\,,
\end{align}
which, for our specific theories \eqref{P_n}, reduces to the condition
\begin{align}
\label{D_cg_1}
    \frac{\cDis}{\cCo}=A_1 =\frac{\alpha}{2}X^{\nth-1}\,.
\end{align}

One representative of this disformal family is specified by, for instance, the additional condition $\cCo=1$, in which case   the  components of the disformed metric \eqref{disformed_metric}, with $\cA=\cB$, read
\begin{align}
    \tilde{\cA}=\cA-q^2 A_1=\cA-\frac{\alpha}{2}q^2X^{\nth-1}\,, \qquad
    \tilde{\cB}=\tilde{\cA}/\left(1-2X A_1\right)=\tilde{\cA}/\left(1-\alpha X^\nth\right)\,.
\end{align}
By making use of \eqref{solAq0} we can deduce the general expression 
\begin{align}
    \tilde{\cA}= 1 -\frac{2\mu}{r}-\frac{\xi_p}{r(p-1)}\int_0^r \frac{\mathrm{d}u}{(1+u^2)^{p-1}}+ \frac{p \, \xi_p}{(p-1)(2p-1)}\frac{1}{(1+r^2)^{p-1}}\,.
\end{align}
One immediately sees that for $p>3/2$, the above integral  converges and $D_\infty$ introduced in \eqref{disformed_infinity} vanishes, which implies that the disformed metric is asympotically flat. Interestingly, this remark  resonates with the fact that the scalar charge \eqref{charge}  converges only for $\nth>3/2$, too.

It is  also worth noticing that the condition \eqref{CondDis}  imposes
an upper bound on the charge of the solution, more precisely:
\begin{eqnarray}
\label{Condxip}
    1 - \alpha \, X^p>0 \; \Longrightarrow \; \xi_\nth < 2\nth-1 \ (\nth\neq 1/2) \quad {\rm or}\quad \alpha^2 q^2 <2 \ (\nth= 1/2) \, ,
\end{eqnarray}
where the parameter $\xi_\nth$ has been defined in \eqref{xi}.

For concreteness, we now discuss the disformal metrics derived from the various black hole solutions discussed in section \ref{sec: examples}.

\subsubsection*{Case $\nth=1/2$}

Starting from the stealth Schwarzschild black hole \eqref{stealth_metric}, we obtain, 
 after performing a disformal transformation  into the $c_g=c$ frame with $C=1$,  the non-homogeneous metric with coefficients
\begin{equation}
    \tilde{\cA}=1-\frac{2M}{r}-\frac{\alpha \abs{q}}{\sqrt{2}}\sqrt{1+{r^2}}\,, \qquad
    \tilde{\cB}^{-1}=\tilde{\cA}^{-1}\left(1-\frac{\alpha \abs{q}}{\sqrt{2(1+{r^2})}}\right) \, ,
    \label{singular}
\end{equation}
which is no longer a stealth solution. 

Interestingly, the condition $\alpha^2 q^2 <2$, which guarantees that the disformal transformation is well defined, also ensures that the solution is free of a naked singularity  as $\tilde{\cB}^{-1}$ does not vanish for any finite $r$. However,  the non homogeneous metric is clearly not asymptotically flat
due to the linear $r$ dependent term in the large $r$ expansion of $\tilde{\cA}$. If $\alpha>0$, we then find, in addition to the BH horizon, a  cosmological horizon, similarly to  a Schwarzschild-de Sitter black hole, and $r$ therefore cannot be arbitrarily large. If $\alpha<0$, the metric coefficient $\tilde \cA$ diverges as $r\rightarrow\infty$, as one would obtain for an Einstein metric with a negative cosmological constant (which is  nonetheless symmetric and regular while it is not the case for the metric here).

\subsubsection*{Case $\nth=1$}
Starting from the homogeneous metric with \eqref{n1}, the  disformed metric in the frame $c_g=c$ and $C=1$ is  given by
\begin{equation}
    \mathrm{d} \tilde{s}^2=-\tilde{\cA}\, \mathrm{d}t_*^2+\tilde{\cA}^{-1}\left(1-\frac{\xi_1}{1+r^2}\right)\mathrm{d} r^2+r^2 \mathrm{d}\Omega^2\,,
\end{equation}
with 
\begin{equation}
 \tilde{\cA}=  1 -3\xi_1-\frac{2 M}{r}-2\xi_1\,\frac{\pi / 2-\arctan r }{r}\,,
 \label{phi_p=1}
\end{equation}
where we have assumed $\xi_1<1$, according to \eqref{Condxip}. 
Similarly to the seed metric,  the disformed metric is only locally asymptotically flat, although with a different solid angle deficit or excess\footnote{One could do away with the solid deficit angle via the disformal transformation $D=-\alpha X^{\nth-1}$ at the price of no longer being in the $c_g=c$ frame. This regularisation evades the problem encountered in \cite{Bakopoulos:2023sdm}.}. The homogeneous and the disformed metrics are thus qualitatively similar. Notice that, in analogy with the discussion below \eqref{n1},  we  require $\xi_1 <1/3$.

\subsubsection*{Case $\nth=2$}
 From the homogeneous metric with \eqref{A_n2}, we obtain another asymptotically flat solution described by the metric
\begin{equation}
    \mathrm{d} \tilde{s}^2=-\tilde{\cA}\, \mathrm{d}t_\star^2+\tilde{\cA}^{-1}\left(1-\frac{\xi_2}{3(1+r^2)^2}\right) \mathrm{d} r ^2+r^2 d\Omega^2\,,
\end{equation}
with
\begin{equation}
 \tilde{\cA}=1-\frac{2 M}{r}+\xi_2\left(\frac{\pi / 2-\arctan r}{r }\right)+\frac{2\xi_2}{3}\frac{1}{1+r^2}  \,,
 \label{An2_disformed}
\end{equation}
where now $\xi_2<3$, following  \eqref{Condxip}.
As we have already pointed out, the disformed solution is regular at the origin, as the seed metric, when  \eqref{Mreg2} is verified.

Interestingly, the horizon $\tilde{r}_h$ of $\tilde{\cA}$, if it exists, is displaced with respect to the horizon $r_h$ (if it exists) of the original metric. 
By comparing \eqref{An2_disformed} with \eqref{A_n2}, one sees that only the last term differs, with a weight $2\xi_2/3$ instead of $\xi_2$. Consequently, we find that  the disformed horizon is larger, i.e. $\tilde{r}_h>r_h$, if $\xi_2>0$ and, conversely, smaller if $\xi_2<0$. This shift has also an impact for regular solutions: in some finite range of masses, the regular solution in the original frame is solitonic whereas the associated disformed solution is a black hole. We will discuss this in more detail in the next section where we consider the effective metric seen by the axial perturbations. Indeed, this effective metric turns out to belong to the $c_g=c$ frame family too, although with $C\neq 1$, so it is conformally related to the above metric and possesses the same horizons.

\subsubsection*{Case $\nth=5/2$}
The case $\nth=5/2$ is very similar to the previous one. The disformal transformation leads to the asymptotically flat non-homogenous black hole defined by the metric
\begin{equation}
    \mathrm{d} \tilde{s}^2=-\tilde{\cA}\, \mathrm{d}t_\star^2+ \tilde{\cA}^{-1}\left(1-\frac{\xi_{_{5/2}}}{4(1+r^2)^{5/2}}\right)\mathrm{d} r^2+r^2 \mathrm{d}\Omega^2\,,
\end{equation}
with
\begin{equation}
\label{An52_disformed}
 \tilde\cA=1-\frac{2 M}{r}+\xi_{_{5/2}}\left(\frac{2}{3r} - \frac{3+8r^2}{12(1+r^2)^{3/2}}\right)\,.
\end{equation}

\subsection{Solutions in the Horndeski frame}
As mentioned earlier, another frame of interest is the Horndeski frame,  in which the disformed theory satisfies the conditions \eqref{Ai_Horndeski}. As already mentioned, such a frame is particularly interesting as the equations of motion are manifestly second order. 
Since the theories we start with belong to the beyond Horndeski class, a purely disformal transformation (i.e. with $C=1$) is sufficient to reach a Horndeski frame. The disformal coefficient  $D(X)$ can be determined by writing the condition
\begin{align}
\tilde A_1=- \tilde \Ftwo_{\tilde{X}}\,,
\end{align}
which characterises Horndeski theories.
Substituting into the above relation the expressions \eqref{disformed_F2} and \eqref{disformed_A1}, with $\cCo=1$, one obtains a differential equation for $\cDis$. In general, such an equation cannot be integrated explicitly and the corresponding Horndeski theory remains only implicitly defined. In our case, however, it turns out that this differential equation can be solved explicitly, yielding\footnote{In the expressions for $D$, we ignore a possible integration constant, associated with the freedom of performing an $X$-independent disformal transformation, which would simply lead to another Horndeski theory. This is due to the fact that the Horndeski class is stable under disformal transformations where $C$ and $D$ do not depend on $X$, as shown in \cite{Bettoni:2013diz}.}
\begin{align}
\label{D_Horndeski}
    \cDis(X)=\frac{(2\nth-1)\alpha}{2(\nth-1)} X^{\nth-1}\quad (\text{for} \; \nth\neq 1)\,,\qquad \cDis(X)=\frac12\alpha \ln(X)\quad (\text{for} \; \nth=1)\,.
\end{align}
By inverting \eqref{disformed_X} after substitution of the above expression for $D$, one can in principle obtain $X$ in terms of $\tilde{X}$ and therefore, using \eqref{disformed_F2} and \eqref{disformed_A1}, get $\tilde{F}$ and $\tilde{A}_1$ as functions of $\tilde{X}$. In practice, this is easily feasible for $p=2$, as \eqref{disformed_X} leads to a quadratic equation for $X$. There are in fact two possibilities, depending on the sign of $\alpha$. The corresponding expressions for $\tilde{F}$ and $\tilde{A}_1$, which define explicitly  the new theory,  are given in Appendix~\ref{App:Horndeski}.

For $\nth=1/2$, the function $D$ in \eqref{D_Horndeski} vanishes, which is consistent with the fact that the original theory is already in the Horndeski sub-class.
For $\nth \neq 1$, if we substitute the first expression above   and $\cA=\cB$ into \eqref{disformed_metric}, we  get  the new metric components
\begin{align}
    \tilde{\cA}=\cA-\frac{(2\nth-1)}{2(\nth-1)}\alpha q^2X^{\nth-1}\,, \qquad
    \tilde{\cB}=\tilde{\cA}/\left(1-\frac{2\nth-1}{\nth-1}\, \alpha X^\nth\right)\qquad (\text{for} \; \nth\neq 1)\,.
\end{align}
By making use of \eqref{solAq0} we can deduce the general expression 
\begin{align}
    \tilde{\cA}= 1 -\frac{2\mu}{r}-\frac{\xi_p}{r(p-1)}\int_0^r \frac{\mathrm{d}u}{(1+u^2)^{p-1}} \qquad (\text{for} \; \nth\neq 1)\,.
\end{align}
 We immediately infer the asymptotic properties of the disformed metric from our previous discussion: for $p >3/2$, the integral converges and  $D_\infty=0$, hence  the seed and disformed metrics behave similarly at spatial infinity, while at the origin (when $r \rightarrow 0$), we recover \eqref{cAatr0}.

By contrast, the case $\nth=1$ in the Horndeski frame is singular due to the logarithmic behavior of $D$ as $r$ goes to infinity. Again, we note a related behaviour in between well defined disformed metrics and  well defined scalar charge \eqref{charge} for $\nth>3/2$.

To conclude this section, let us give the expressions of the Horndeski frame metrics in our two  well-behaved cases. For $\nth=2$, the Horndeski metric components are given by
\begin{equation}
 \tilde{\cA}=1-\frac{2 M}{r}+\xi_2\left(\frac{\pi / 2-\arctan r }{r }\right)\, ,\qquad \tilde{\cB}=\tilde{\cA}/\left(1-\frac{\xi_2}{(1+r^2)^2} \right) \, ,
 \label{metric_n2_Horndeski}
\end{equation}
while, in the  case $\nth=5/2$, we find 
\begin{equation}
 \tilde{\cA}=1-\frac{2 M}{r}+\frac23\xi_{_{5/2}}\left(\frac{1}{r}- \frac{1}{\sqrt{1+r^2}}\right),
 \qquad \tilde{\cB}=\tilde{\cA}/\left(1-\frac{2  \, \xi_{_{5/2}}}{3(1+r^2)^{5/2}}  \right)\,.
 \label{metric_n21_Horndeski}
\end{equation}
They are very similar to their counterparts \eqref{An2_disformed} and \eqref{An52_disformed} in the $c_g=c$ frame. 

\section{Axial perturbations}
\label{Section_axial}
We now consider the  axial linear  perturbations $h_{\mu\nu}$ about the background metric ${g}_{\mu\nu}^{\rm bgd}$, introduced in \eqref{background_metric}, defined by
\begin{equation}
    g_{\mu\nu}=g_{\mu\nu}^{\rm bgd} +h_{\mu\nu}\,,
\end{equation}
while the scalar field remains unperturbed.
It is convenient to work in the so-called Regge-Wheeler gauge \cite{Regge:1957td} where axial perturbations are described by the following non-vanishing components of the perturbations:
\begin{align}
	 & h_{t\theta} = \frac{1}{\sin\theta}  \sum_{\ell, m} h_0^{\ell m}(t,r) \partial_{\varphi} {Y_{\ell m}}(\theta,\varphi), \qquad
	h_{t\varphi} = - \sin\theta  \sum_{\ell, m} h_0^{\ell m}(t,r) \partial_{\theta} {Y_{\ell m}}(\theta,\varphi), \nonumber         \\
	 & h_{r\theta} =  \frac{1}{\sin\theta}  \sum_{\ell, m} h_1^{\ell m}(t,r)\partial_{\varphi}{Y_{\ell m}}(\theta,\varphi), \qquad
	h_{r\varphi} = - \sin\theta \sum_{\ell, m} h_1^{\ell m}(t,r)  \partial_\theta {Y_{\ell m}}(\theta,\varphi), \label{eq:odd-perttext}
\end{align}
using an expansion in spherical harmonics ${Y_{\ell m}}(\theta,\varphi)$, which reflects the spherical symmetry of the background. In the following, since perturbations with different values of $\ell$ and $m$ do not couple at  linear level, we drop the indices $\ell$ and $m$ to shorten the equations. 
Moreover,  we consider only $\ell \geq 2$ since axial perturbations contain no monopole ($\ell=0$) nor dipole ($\ell=1$) contributions.

\subsection{Equations of motion: first order system}
We follow the method introduced and developed in \cite{Langlois:2021aji, Langlois:2021xzq, Langlois:2022ulw} to study the dynamics of static and spherically symmetric black hole perturbations in scalar-tensor theories. 
It is convenient to introduce a new time coordinate,
\begin{eqnarray}
\label{t_star_eff}
t_* = t - \int \mathrm{d}r \,  \Psi(r) \,,
\end{eqnarray}
where, in general,   $\Psi$ is defined by \cite{Langlois:2022ulw}
\begin{equation}
  \Psi = q \frac{\psi' A_1}{\mathcal{F}}\,, \qquad  \mathcal{F} =  \cA (\Ftwo+2X A_1) - q^2 A_1  \,.
	 \end{equation}
For the theories \eqref{P_n}, the expression for $\mathcal{F}$ reduces to $\mathcal{F}=\cA - q^2 A_1$. It thus turns out that the above definition of $t_*$ coincides with the definition \eqref{t_star} when $\cDis/\cCo=A_1$, for reasons that will become clear below.

We then decompose the two variables $h_0$ and $h_1$,  defined in \eqref{eq:odd-perttext}, in Fourier modes,
\begin{eqnarray}
    h_a(r,t) = \int \mathrm{d}\omega \, e^{-i \omega t_*} \, h_a(r,\omega)  \, , \quad \text{for} \; a=0,1 \,,
\end{eqnarray}
using the same notation for $h_a$ in real space and in the frequency domain for simplicity. From now on, we work in the frequency domain exclusively. 
As shown in \cite{Langlois:2022ulw}, the equations of motion for the two perturbations $h_0$ and $h_1$ can be written in the first-order form
\begin{equation}
\label{EOM_1st_order}
	\dv{Y}{r}= \mathbf{M} \, Y \, , \quad \text{with} \quad \mathbf{M}=
	\begin{pmatrix}
		 {2}/{r} & -i\omega^2 + {2 i \lambda \Phi}/{r^2} \\
		- i \Gamma & \Delta 
	 \end{pmatrix} \, ,
\end{equation}
where  the vector $Y$ is given by
\begin{equation}
Y = \left( \begin{array}{c} Y_1 \\ Y_2 \end{array}\right) \qquad \text{with} \qquad Y_1=h_0\, , \quad
	\omega Y_2 = h_1 + \Psi h_0  \,.
\end{equation}
All the coefficients appearing in \eqref{EOM_1st_order}  are defined, for a generic DHOST theory,  in terms of the action functions, the metric and scalar background solutions, by the relations~\cite{Langlois:2022ulw}
\begin{eqnarray}
&&\Phi =\frac{\mathcal{F}}{\Ftwo +2X A_1} \,, \qquad \Delta = - \frac{\mathcal{F} '}{\mathcal{F} } - \frac{\cB'}{2\cB} + \frac{\cA'}{2\cA}\,,\nonumber \\
&&\Gamma = \Psi^2 + \frac{1}{\cA\cB\mathcal{F}} ( {q^2 A_1} + \cA {\Ftwo} )=\frac{\cA \Ftwo (\Ftwo+2X A_1)}{\cB \mathcal{F}^2 } \,, \label{PhiGammaDeltaPhi}
\end{eqnarray}
where the second expression for $\Gamma$ is derived upon using $
    \cB \, \psi^{\prime 2}={q^2}/{\cA}-2X$, 
which is a direct consequence of the definition  of $X$. 

For the theories \eqref{P_n} and the homogeneous solutions that we are considering in this paper, the expressions of these coefficients simplify and reduce to
\begin{align}
\label{coeffs_here}
\Phi =\mathcal{F} =  \cA - q^2 A_1\,, \qquad\Delta = - \frac{\Phi'}{\Phi} \,,\qquad
	\Gamma =\frac{\Ftwo}{\Phi^2 } \,,
\end{align}
since $\Ftwo+2X A_1=1$ and $\cB=\cA$.

\subsection{Effective metric for axial perturbations}
As shown in \cite{Langlois:2022ulw} (see also \cite{Tomikawa:2021pca}),  there is a correspondence, in {\it quadratic} DHOST theories\footnote{This result has been extended to cubic DHOST theories in \cite{Noui:2023ksf}}, between the dynamics
of axial perturbations about the background metric \eqref{background_metric} and the standard GR dynamics of axial perturbations  in a {\it different} (although still static and spherically symmetric) metric $\hat{g}_{\mu\nu}$, which we call the effective metric,   given by
\begin{eqnarray}
\label{effectivemetric_DHOST}
 \dd{\hat s}^2 &=& \abs{ \mathcal{F}}  \sqrt{ \frac{\Gamma \cB}{\cA}}\left( 
 - \Phi \dd{t_*}\!\!\!^2 + \Gamma \Phi \dd{r}^2 + r^2 \dd{\Omega}^2 \right) \, .
 \end{eqnarray}
In our case, using the coefficients \eqref{coeffs_here}, the effective metric reduces to 
\begin{equation}
\label{effectivemetric}
\mathrm{d} \hat{s}^2=\sqrt{\Ftwo}\left(-\Phi \, \mathrm{d} t_*^2+\frac{\Ftwo}{\Phi} \, \mathrm{d} r^2+r^2 \mathrm{d} \Omega^2\right) \, \equiv \; -\hat{\cA}(r) \, \mathrm{d} t_*^2+\frac{\mathrm{d} r^2}{\hat{\cB}(r)} +\hat{\cC}(r) \, \mathrm{d} \Omega^2\,.
\end{equation}

As discussed in \cite{Langlois:2022ulw}, this effective metric coincides with the disformed metric obtained when going to  the Einstein frame, characterised by
\begin{equation}
\label{Einstein_frame}
    \hat{\Ftwo}=1\,, \qquad \hat{A_1}=0\,,
\end{equation}
 where axial perturbations behave exactly as in GR.
Indeed, from \eqref{disformed_F2} and \eqref{disformed_A1}, one sees that these conditions  are satisfied 
by choosing\footnote{We assume here that $\hat \Ftwo>0$, i.e. $\Ftwo>0$, otherwise the effective Planck mass squared is negative, leading to a theory plagued with ghost instabilities.}
\begin{eqnarray}
\cCo \, =  \, \sqrt{\Ftwo(\Ftwo +2 X A_1)}=  \, \sqrt{\Ftwo} \,, \qquad \cDis \,  = \, \frac{ \Ftwo A_1}{\sqrt{\Ftwo(\Ftwo +2 X A_1)}}=\sqrt{\Ftwo}\,  A_1 \,,
\label{coeffs_disformal}
\end{eqnarray}
where we have used, once again,  the property $\Ftwo+2X A_1=1$. The corresponding disformed  metric is thus given by
\begin{eqnarray}
	\label{alphaquadratic}
	\hat{g}_{\mu\nu} =   \sqrt{\Ftwo} \left(  g_{\mu\nu} +  A_1\, \phi_\mu \phi_\nu \right) \,,
\end{eqnarray}   
which can be checked to coincide with \eqref{effectivemetric}.

Importantly, the Einstein frame \eqref{Einstein_frame} belongs to the  $c_g=c$ frame family, because of the second condition in \eqref{Einstein_frame}. This implies that the Einstein frame is conformally related to the $c_g=c$ frame with $C=1$ that we introduced 
in subsection \ref{sectioncgc}. Since a conformal transformation preserves the causal structure of the metric, the whole discussion of subsection \ref{sectioncgc} remains valid for the Einstein frame, with the only difference that the explicit expressions for $\tilde{\cA}$ and $\tilde{\cB}^{-1}$ given there must now be multiplied by $\sqrt{\Ftwo}$.

In summary, axial gravitational waves seem to propagate in an effective space-time whose metric differs from the background metric, the latter defining  how light propagates. This means, in particular, that gravitons and photons ``see'' different horizons, when both background and effective metrics are black holes. In the following, we will denote by $r_g$  the horizon of the effective metric, or gravitational horizon,  and by $r_l$ the horizon of the background metric, or luminous  horizon.
In the limit where $\xi_p$ is very small both horizons are very close to the Schwarzschild horizon $r_S=2M$, with
\begin{equation}
\label{horizons}
r_l\simeq r_S+2\xi_p \, \check\Xi_p(2M)\,, \qquad r_g-r_l\simeq \frac{2M \, \xi_p}{(2p-1)(1+4M^2)^{p-1}}\,,
\end{equation}
where we have used $q^2A_1(X(r))=\xi_p(1+r^2)^{1-p}/(2p-1)$.

\begin{figure}[h]
\begin{center}
\includegraphics[width=5.5cm]{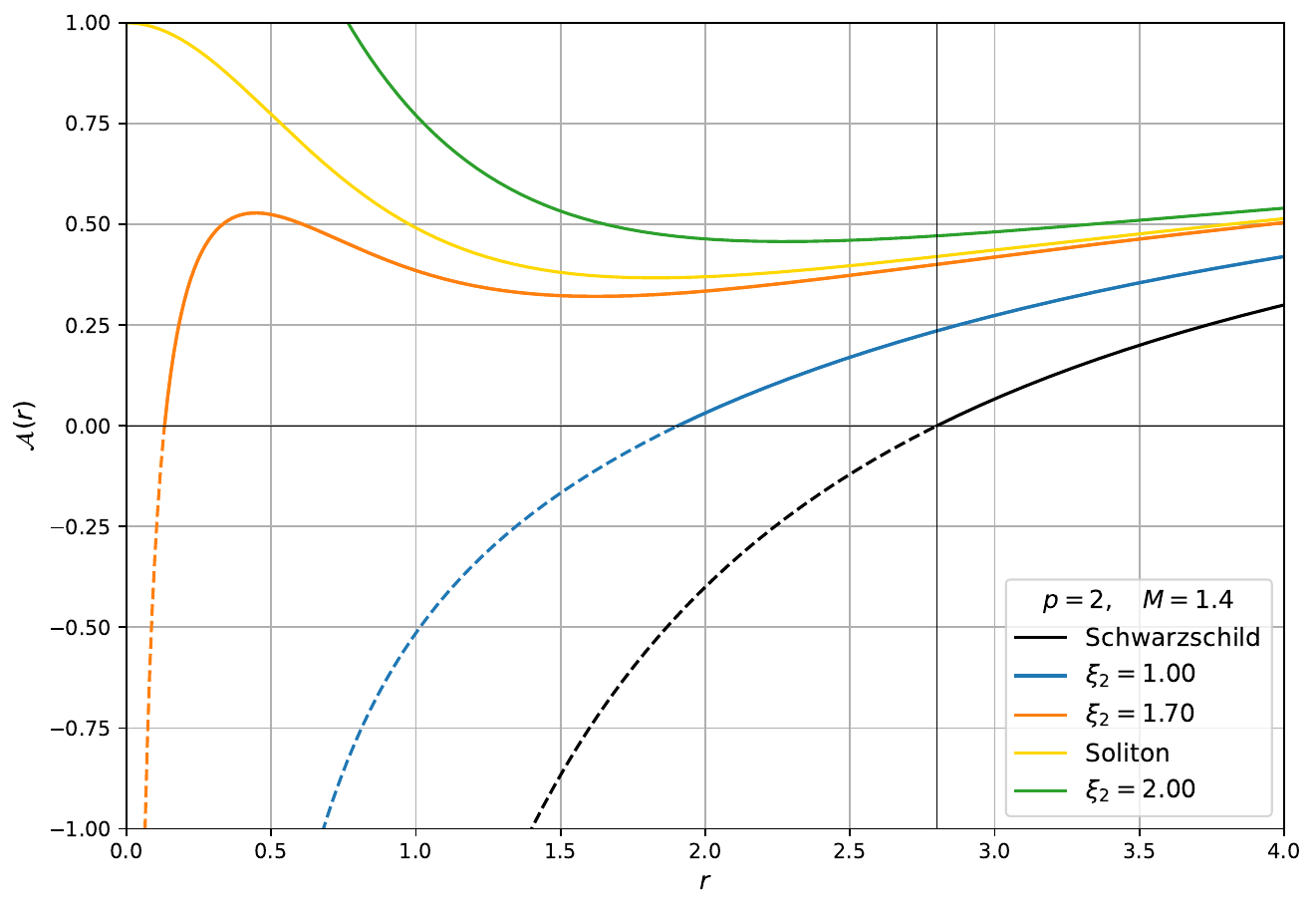}
\includegraphics[width=5.5cm]{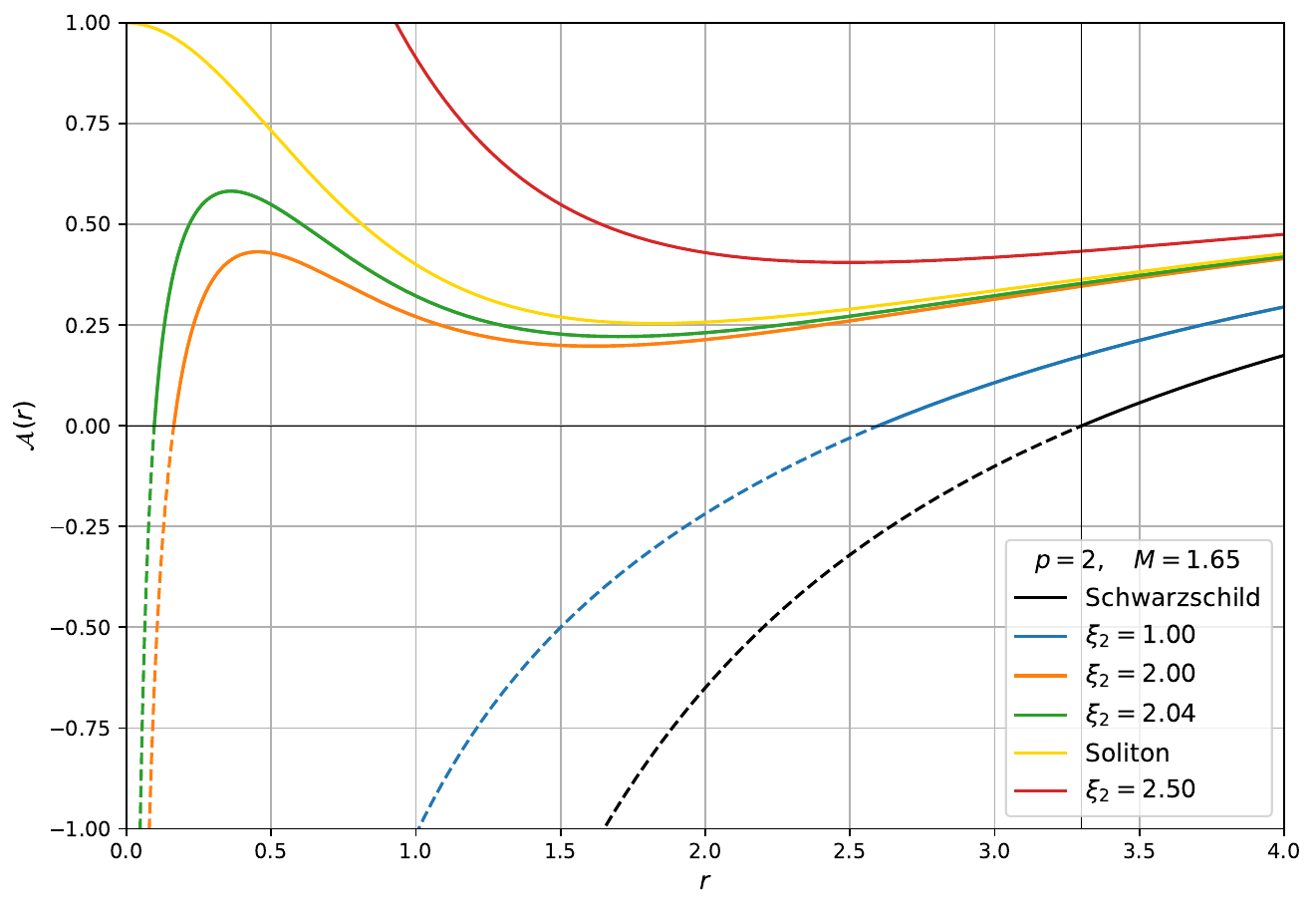}
\includegraphics[width=5.5cm]{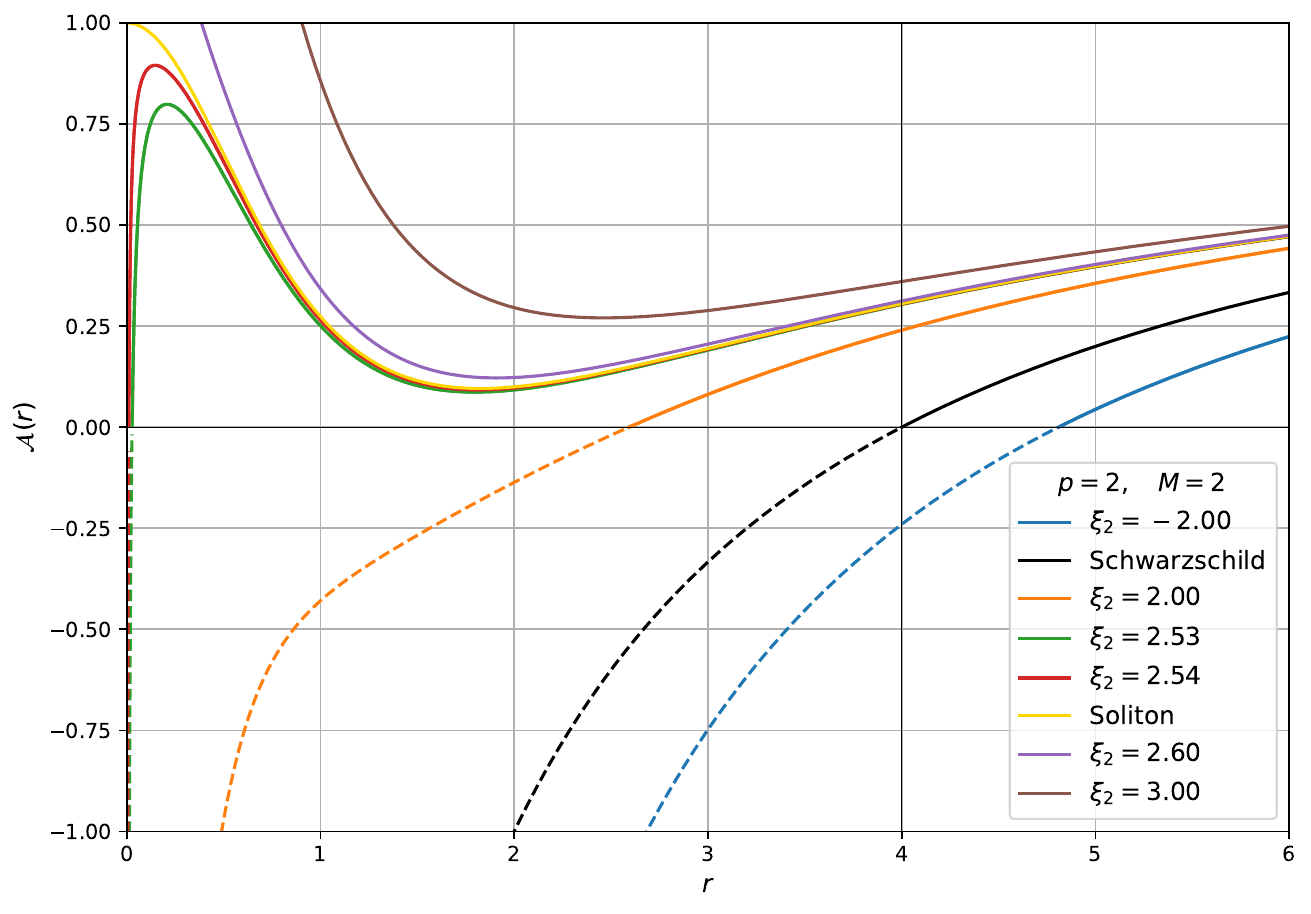}

\vspace{0.3cm}

\includegraphics[width=5.5cm]{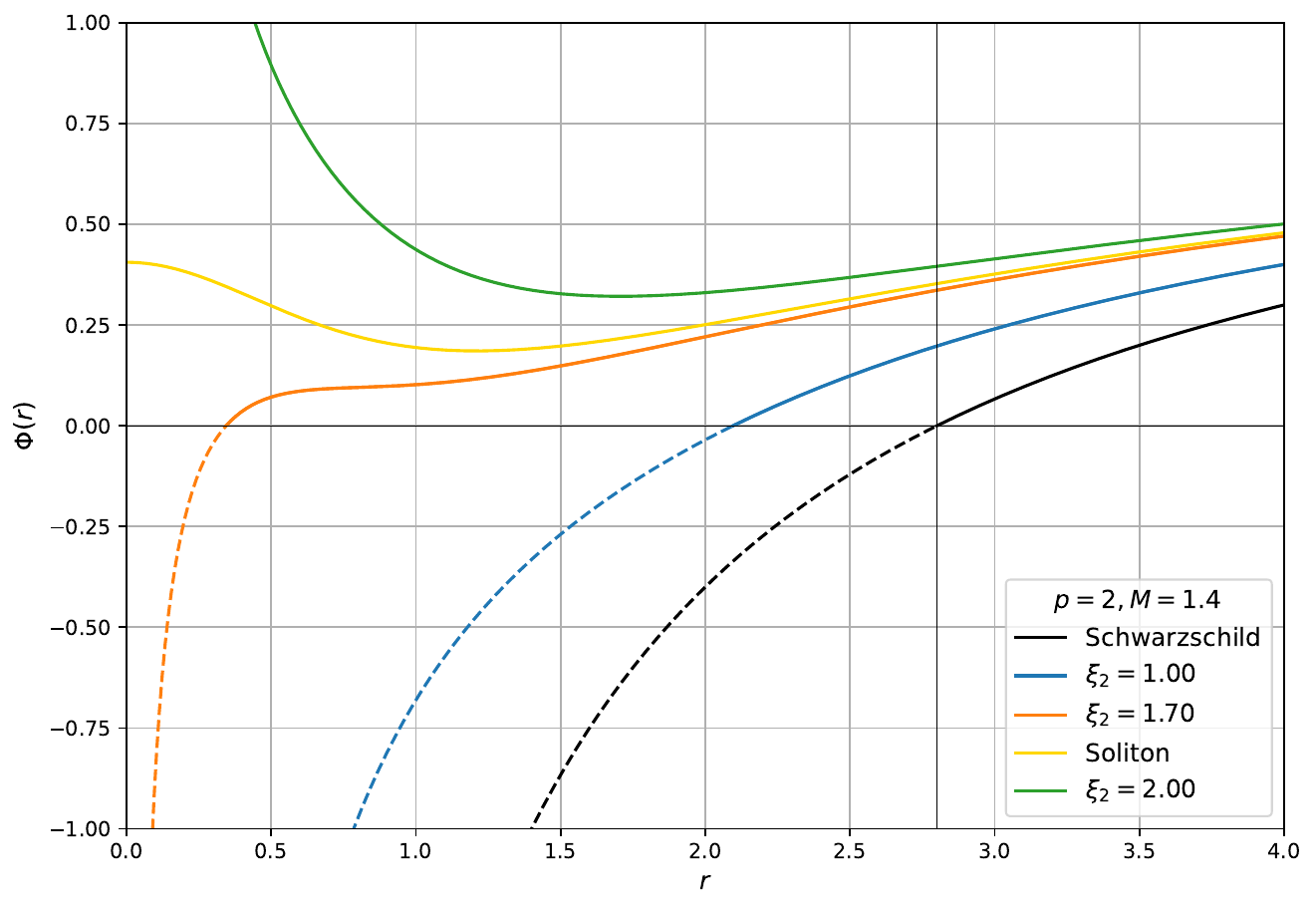}
\includegraphics[width=5.5cm]{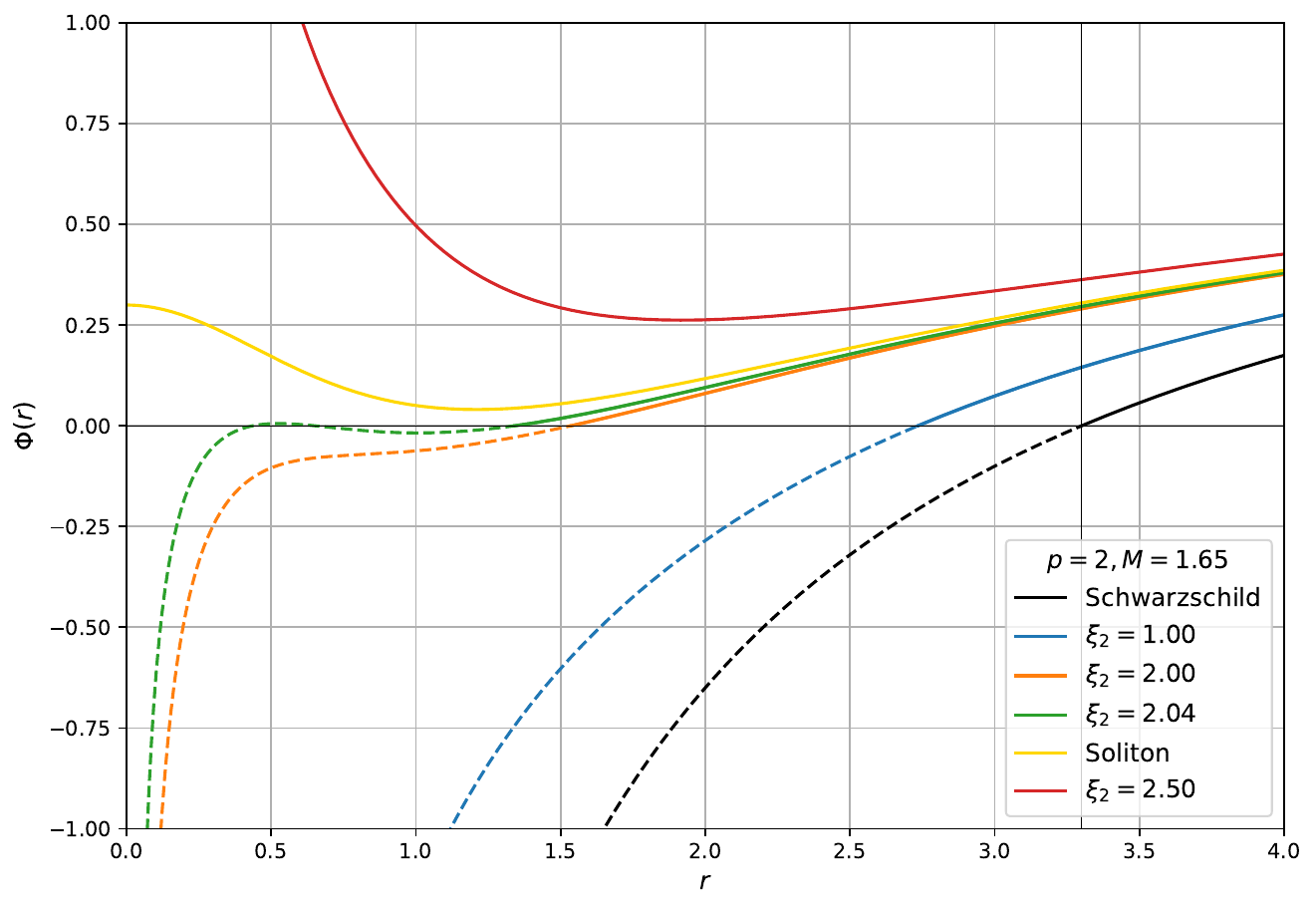}
\includegraphics[width=5.5cm]{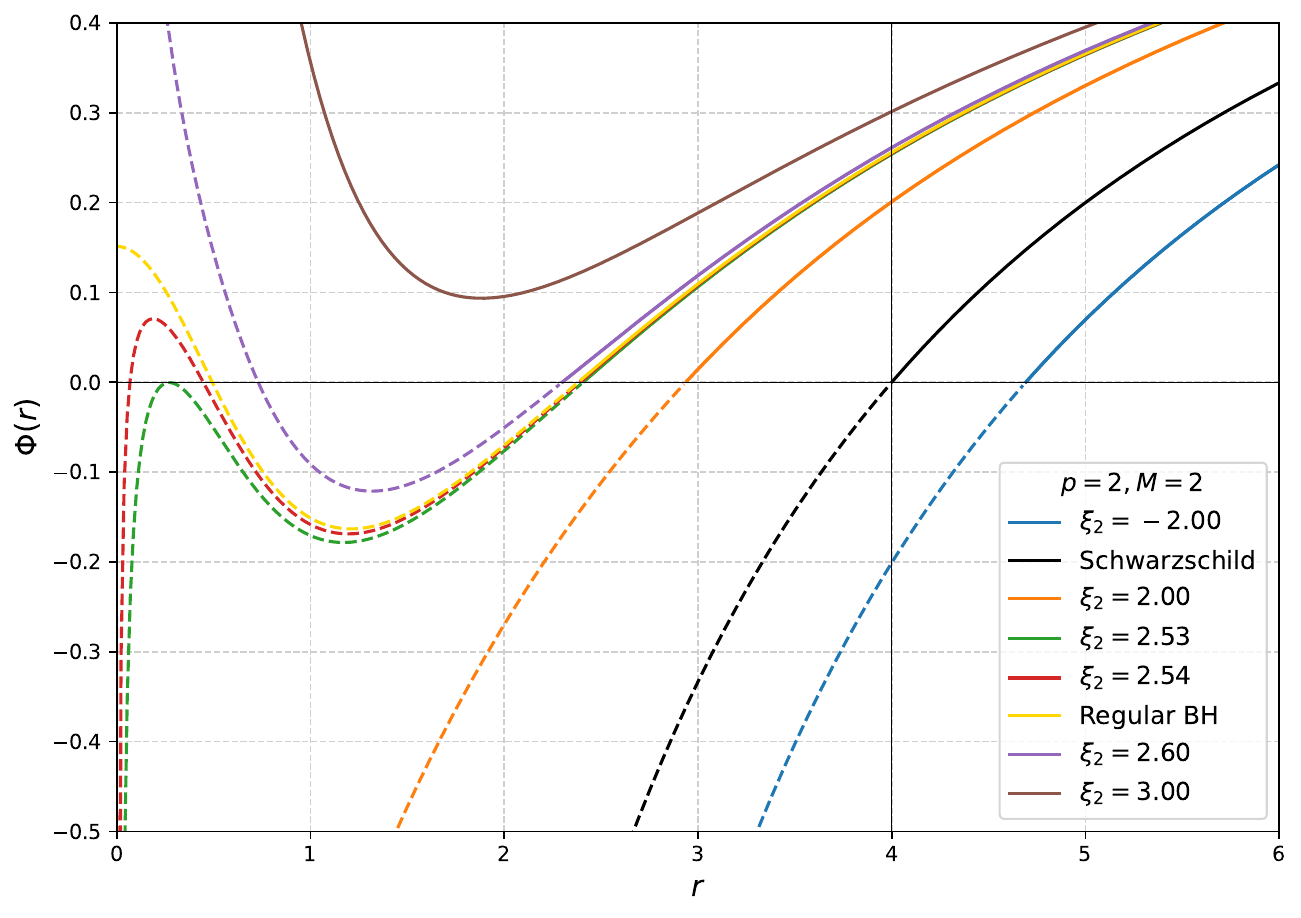}
\end{center}
\caption{Plots of the background metric coefficient $\cA(r)$ (top) and  of the effective metric coefficient $\Phi(r)$ (bottom) in the theory $\nth=2$, for three value  of the mass  ($M=1.4,1.65, 2$) and several values of the  parameter $\xi_2$, ranging from  $-2$ to the upper bound $\xi_{2}=3$. The Schwarzschild  ($\xi_2=0$) and regular  solutions correspond to the black and yellow curves, respectively. The dashed parts of the curves indicate  the regions hidden behind the horizon.}
\label{fig:n=2_intermediate}
\end{figure}

\begin{figure}[!h]
\begin{center}
\includegraphics[width=8.4cm]{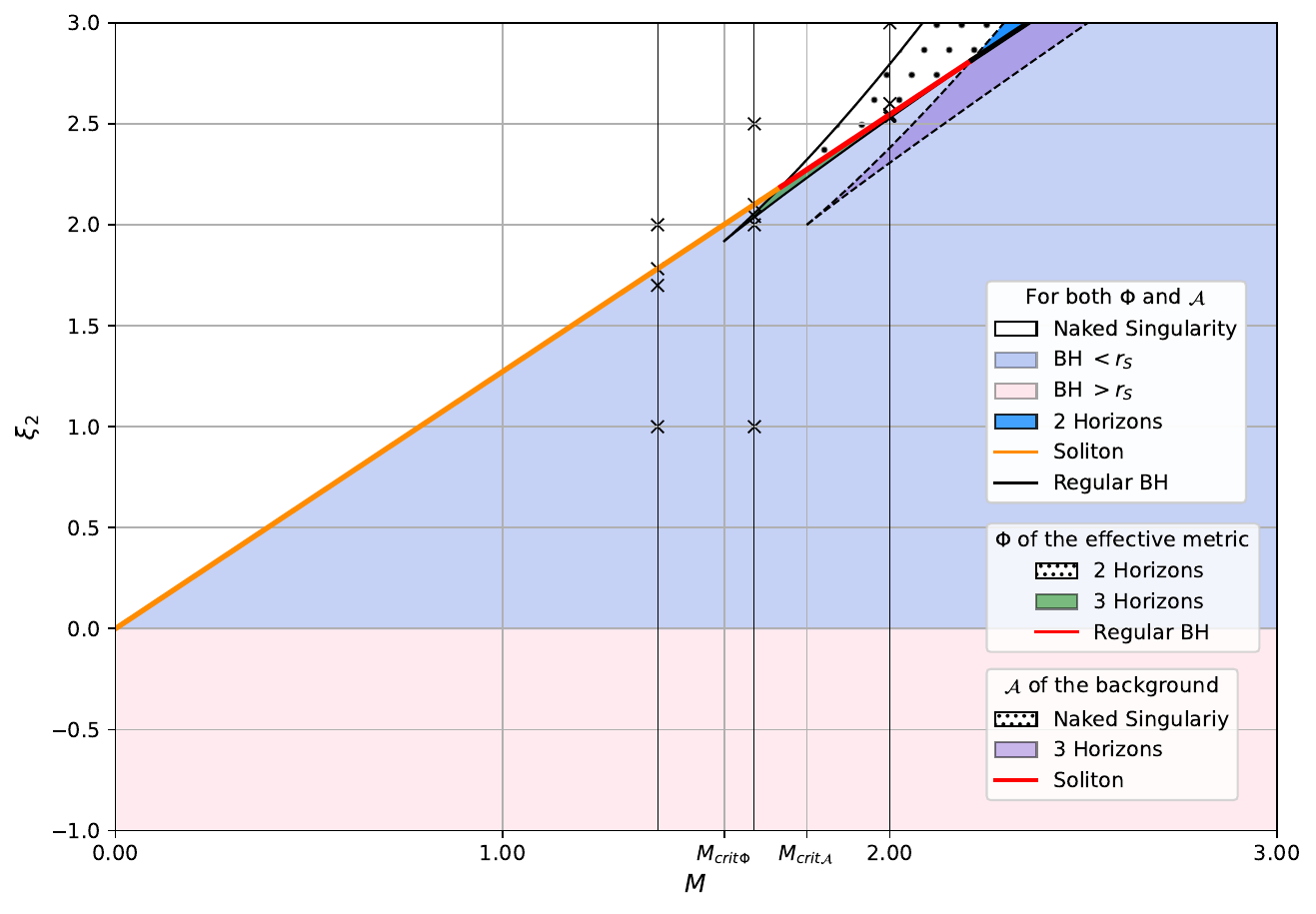}
\includegraphics[width=8.4cm]{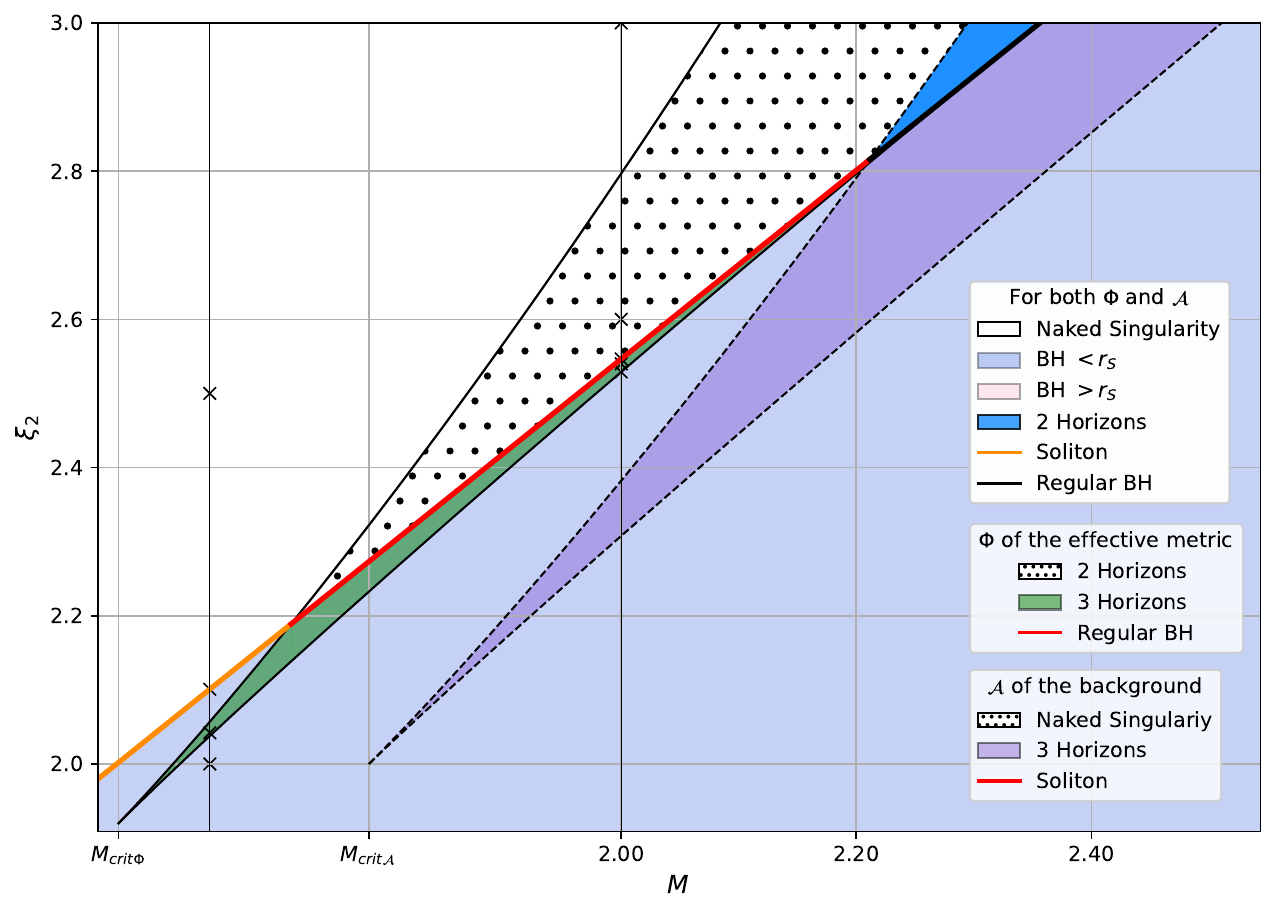}
\end{center}
\caption{
Left: ``Phase diagram'' displaying  the background and effective metrics, for a wide range of parameters $M$ and $\xi_2$, in theories with $\nth=2$. 
$M_{{\rm crit} \Phi}\simeq 1.572$ ($M_{{\rm crit} \mathcal{A}}\simeq 1.719$) is the critical mass beyond which multiple horizons appear for the effective (background) metric. 
The crosses on the black vertical lines correspond to the parameters chosen for the plots in Fig.~\ref{fig:n=2_intermediate}.
Right: focus on the most intricate region of the phase diagram.}
\label{fig:Phase_diagram}
\end{figure}

Let us illustrate in the case $p=2$ the various configurations for the background and effective metrics, including the presence or not of horizons.
In Fig.~\ref{fig:n=2_intermediate}, we have plotted the background and effective metric components for certain values of the mass $M$ and of the  parameter $\xi_2$. 
More generally, all possible situations are summarised in the ``phase diagram'' of Fig.~\ref{fig:Phase_diagram}. The particular cases plotted in Fig.~\ref{fig:n=2_intermediate}  are indicated by crosses in the phase diagram. The boundaries separating the different regions are explicitly computed in Appendix \ref{App:phase_diagram}.
Let us discuss the most important properties illustrated in these figures.

When  $\xi_2<0$,  corresponding to the pink shaded region in the phase diagram,  we find that both the background and effective metrics possess a horizon, and their ordering, for the same mass,  is the following: $r_S<r_g<r_l$. For small values of $\xi_2$, the relative shift between the gravitational  and luminous horizons is given by \eqref{horizons} with $p=2$.

For $\xi_2>0$, the possibilities are much more diverse. The particular case where $\xi_2=4M/\pi$, in which case both metrics are regular at the origin according to \eqref{Mreg2}, corresponds to a straight line in the phase diagram (and yellow curves in Fig.~\ref{fig:n=2_intermediate}).
Along this line, as $M$ increases, one can find three distinct situations: for $M<M^{\rm reg,I}_2$, both metrics describe solitons, as there is no horizon. For $M^{\rm reg,I}_2<M<M^{\rm reg,II}_2$, the background metric is still a soliton while the effective metric describes a regular black hole. Finally, for  $M>M_2^{\rm reg,II}$, both metrics correspond to regular black holes. The values of  $M_2^{\rm reg,I}$ and $M_2^{\rm reg,II}$ are given in Appendix \ref{App:phase_diagram}.

Above the regularity line in the phase diagram, one usually  finds a naked singularity for the background solution, together with a naked singularity (white region) or a BH (white dotted region) for the effective metric, except in a small region (blue region) where both metrics are BHs.   
Below the regularity line, both metrics describe BHs, with an odd number of horizons: a single horizon (light blue region) in most cases, but also three horizons for either the background metric (purple region) or the effective metric (green region).

For $\xi_2>0$, the outer horizons, when they exist, satisfy the inequalities $r_l<r_g<r_S$, i.e. the opposite of the inequality for a negative $\xi_2$.

\subsection{Schr\"odinger-like equation}

As in GR, the equations of motion for the axial modes propagating in a metric of the form \eqref{effectivemetric} can be reformulated as a Schr\"odinger-like equation (in the frequency domain), of the form \cite{Langlois:2022ulw}
\begin{eqnarray}
\label{schro}
    -\frac{\dd^2 {\cal Y}}{\dd \hat r_*^2} + V_{\ell} \, {\cal Y} \; = \; \omega^2 \, {\cal Y} \, ,
\end{eqnarray}
where 
\begin{equation}
    {\cal Y}=\frac{\Phi}{r \Ftwo^{1/4}}Y_2=\frac{\Phi}{r \Ftwo^{1/4}\omega}(h_1 + \Psi h_0)
\end{equation}
and $\hat r_*$ is the tortoise coordinate associated with the {\it effective} metric, i.e. $\dd \hat r_*=(\sqrt{\Ftwo}/\Phi) \dd r$. The explicit expression of the  potential is given by
\begin{equation}
V_{\ell}=(\ell(\ell+1)-2) \frac{\hat{\cA}}{\hat{\cC}}+\frac{1}{2} \frac{\hat{\mathcal{D}}^2 \hat{\cC}^{\prime 2}}{\hat{\cC}}-\frac{1}{2} \hat{\mathcal{D}}\left(\hat{\cC}^{\prime} \hat{\mathcal{D}}\right)^{\prime}\,, \quad \hat{\mathcal{D}}\equiv \sqrt{\hat{\cA}\hat{\cB}/\hat{\cC}}\,,
\end{equation}
where a prime denotes a derivative with respect to $r$. 
Note that in GR,  where the effective metric coincides with the background Schwarzschild metric, i.e. $\hat\cA=\hat\cB=\cA_S\equiv 1-2M/r$ and $\hat{\cC}=r^2$, the above expression reduces to
\begin{align}
\label{potentialGR}
V_{\ell}(r)=&\frac{\ell^2+\ell-2}{r^2}\cA_S-\frac{1}{r}\cA_S\cA_S'+\frac{2}{r^2}\cA_S^2\,,
\end{align}
which yields the well-known Regge-Wheeler potential \cite{Regge:1957td}.

Substituting the expression of the effective metric in terms of 
$\Ftwo$ and $\Phi$, given in \eqref{effectivemetric}, one can write the  potential in a form very similar to the above GR expression, but with the effective component $\Phi$
replacing the background component $\cA$ and more complicated coefficients for the last two terms:
\begin{align}
\label{pot2}
V_{\ell}(r)=&\frac{\ell^2+\ell-2}{r^2}\Phi-\frac{\kappa_1}{r}\Phi\Phi'+\frac{2 \kappa_2}{r^2}\Phi^2\,,
\end{align}
with the functions $\kappa_1$ and $\kappa_2$  given by
\begin{equation}
    \kappa_1=\frac{r \Ftwo' +4  \Ftwo }{4\Ftwo^2} \, ,  \qquad
    \kappa_2=\frac{r^2 \left(7\Ftwo^{\prime 2}-4 \Ftwo\Ftwo''\right)+16r \Ftwo\Ftwo'+32
   \Ftwo^2}{32  \Ftwo^3} \,,
\end{equation}
where $\Ftwo$ is treated here as a function of $r$, defined by $\Ftwo(r)\equiv \Ftwo(X(r))$ and using the explicit expression \eqref {scalar}\footnote{Conversely, one can reexpress $\kappa_1(r)$ and $\kappa_2(r)$ in terms of $X$ by expressing $r$ in terms of $X$ using \eqref{scalar}.
This yields 
$   \kappa_1=\left[1+\alpha(\frac{\nth}{2}-1)X^\nth-\frac{\alpha \nth}{q^2} X^{\nth+1}\right]/(1-\alpha X^\nth)^2$
and a more involved expression for $\kappa_2$.}. At spatial infinity, i.e. $r\to \infty$,  one can check that $\kappa_1=\kappa_2=1$.

\begin{figure}[h]
\begin{center}
\includegraphics[width=8.4cm]{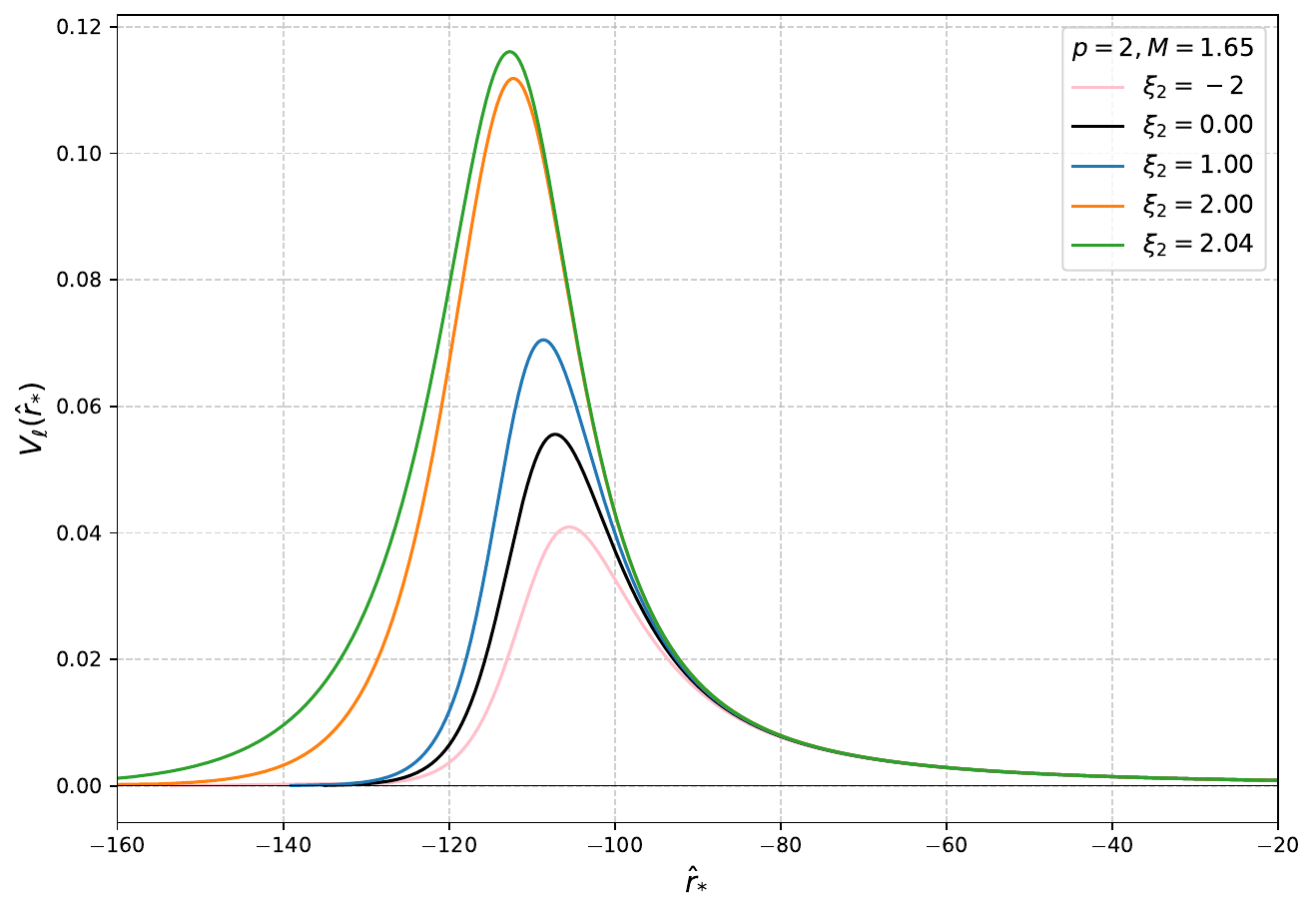}
\includegraphics[width=8.4cm]{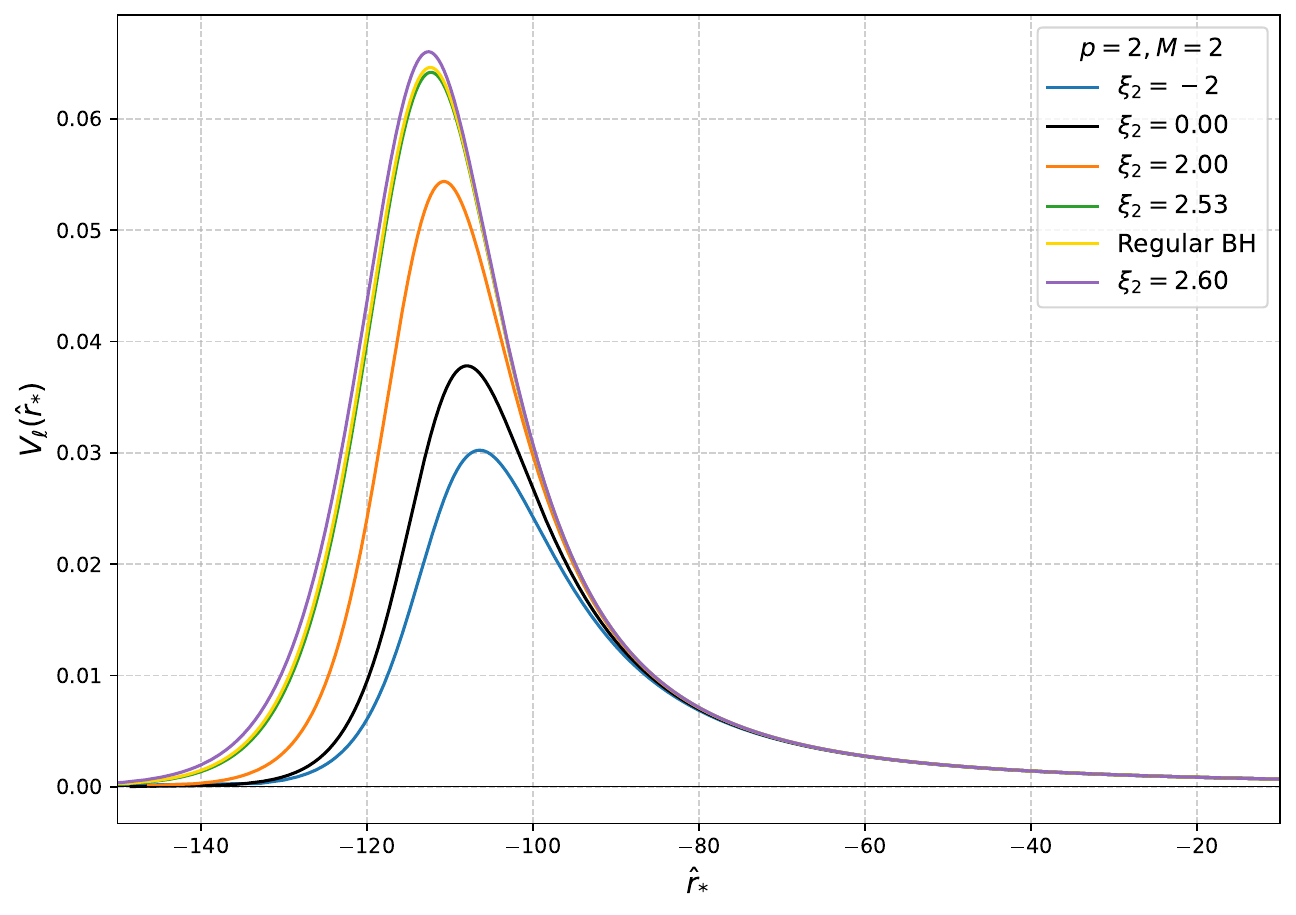}
\end{center}
\caption{Plot of $V_{\ell}(r_{*})$ for $\nth=2$ and $M=1.65$ on the left and $M=2$ in the right plot. We vary the parameter $\xi_{2}$ between $-2$ and $3$, passing by the regular value $4M/\pi $ which corresponds to a soliton in the left plot, while on the right, it describes a regular BH. 
Note that each value of $\xi_2$ corresponds to a distinct tortoise coordinate $\hat{r}_*$. 
 }
\label{fig:potential}
\end{figure}
In Fig.~\ref{fig:potential}, we have plotted the potential \eqref{pot2} as a function of $\hat r_*$ for the theory $\nth=2$ (recall that the definition of $\hat r_*$ depends on $\xi_p$): we show the potential for two different masses, varying the parameter $\xi_2$ in each case. It is worth noting that the potential is proportional to $\Phi$ and therefore vanishes at the gravitational horizon $r_g$, which corresponds to the limit $\hat r_*\to -\infty$. The potentials such as those in Fig.~\ref{fig:potential} are thus associated with the  exterior region $r>r_g$.
We will use these potentials in the next section to compute numerically the associated QNMs.

\section{Estimation of the quasi-normal modes}
\label{Section_QNMs}
In this section, we  provide  a numerical estimation, using the WKB method, of the quasi-normal modes (QNMs) associated with the Schr\"odinger equation \eqref{schro}. We begin with a brief review of the WKB approximation before presenting our results for  several  choices of the potential \eqref{pot2}, related to the effective metrics examined earlier.

\subsection{The WKB method}
There exist many techniques to compute the QNMs associated with a Schrödinger-like equation of the form \eqref{schro} (see e.g. the reviews \cite{Kokkotas:1999bd,Berti:2009kk,Konoplya:2011qq}). Since very few potentials admit analytic expressions for QNMs, most techniques rely on numerical methods\footnote{Among these, a particularly effective method was introduced and developed by Leaver \cite{Leaver:1985ax}, achieving exceptional precision in the computation of QNM spectra for Schwarzschild and Kerr black holes in GR. Leaver’s approach relies on solving a continued fraction equation, a technique that has recently been adapted to compute QNMs in modified gravity \cite{Roussille:2023sdr}. However, this method becomes significantly more challenging to apply when dealing with potentials that are not rational functions of the radial coordinate, which is precisely the situation encountered for the axial perturbations considered in this work.}.
Here, we  consider the well-known WKB method that was first introduced to solve equations in quantum physics, and later adapted and improved for black hole QNMs~\cite{Iyer:1986np,schutz_black_1985}. 
This method, valid when the potential $V$ admits a single maximum  and tends to constants at $\pm \infty$, gives an approximation of the lowest QNMs.

Labelling the QNMs  with an integer  (with $n=0$ for the fundamental mode and $n\geq 1$ for the successive  overtones), the QNM (complex) frequencies are approximately given by  the expression~\cite{Konoplya:2011qq}
\begin{eqnarray}
\label{WKBexpansion}
    \omega_n^2 \simeq  V_{\rm max} - i  \sqrt{-2 V''_{\rm max}}\left(n+\frac{1}{2} + \sum_{i=2}^N \Lambda_i \right) \, ,
\end{eqnarray}
where $V_{\rm max}$ and $V''_{\rm max}$ are the potential and its second derivative evaluated at the radius $\hat r_{*}^{\rm max}$  corresponding to the maximum of the  potential\footnote{At the maximum of the potential, we always have $V''_{\rm max}<0$ and the square root is  thus a real positive number.}, a prime denoting here a derivative with respect to $\hat r_{*}$. The integer $N$ is the order of the approximation and the coefficients $\Lambda_i$  depend on the derivatives of the potential at $\hat r_{*}^{\rm max}$ (their explicit expression can be found in e.g.~\cite{Konoplya:2011qq}).

The expression \eqref{WKBexpansion} is based on a Taylor expansion of the potential around its maximum, truncated at order $N$. Importantly, if $N$ were taken to infinity, the series would become divergent. This reflects the fact that the WKB expansion is not a convergent series, but rather an asymptotic series, where only the first few terms provide an accurate approximation of the QNM frequencies. 
Depending on both the shape of the potential and the particular QNM being considered, there exists an optimal truncation order $N_{\rm opt}$ that minimises the error. For higher values of $N$, the accuracy deteriorates as the series begins to diverge~\cite{hatsuda_2020}. It is known that the WKB approximation is most effective for the fundamental mode and lowest overtones.

\subsection{Numerical  results}
In order to estimate the QNM frequencies for some of the  potentials considered in this work,  we have used the Mathematica notebook \cite{WKBpackage}, which enables us to compute order by order the QNM frequencies for any potential  up to order $N=13$. In practice, we proceed as follows. For any QNM, we calculate the frequencies $\omega_n^{(N)}$ for several truncation orders $N$ in the WKB approximation, using \eqref{WKBexpansion}. We then  compute the convergence parameter 
\begin{equation}
\Delta_n^{(N)}={\left|\omega_{n}^{(N+1)}-\omega_{n}^{(N-1)}\right|} \,,
\end{equation}
and keep the frequency corresponding to the order $N$  with the lowest value of $\Delta_{n}^{(N)}$. 

\medskip

\begin{figure}[h]
\begin{center}
\includegraphics[width=8.4cm]{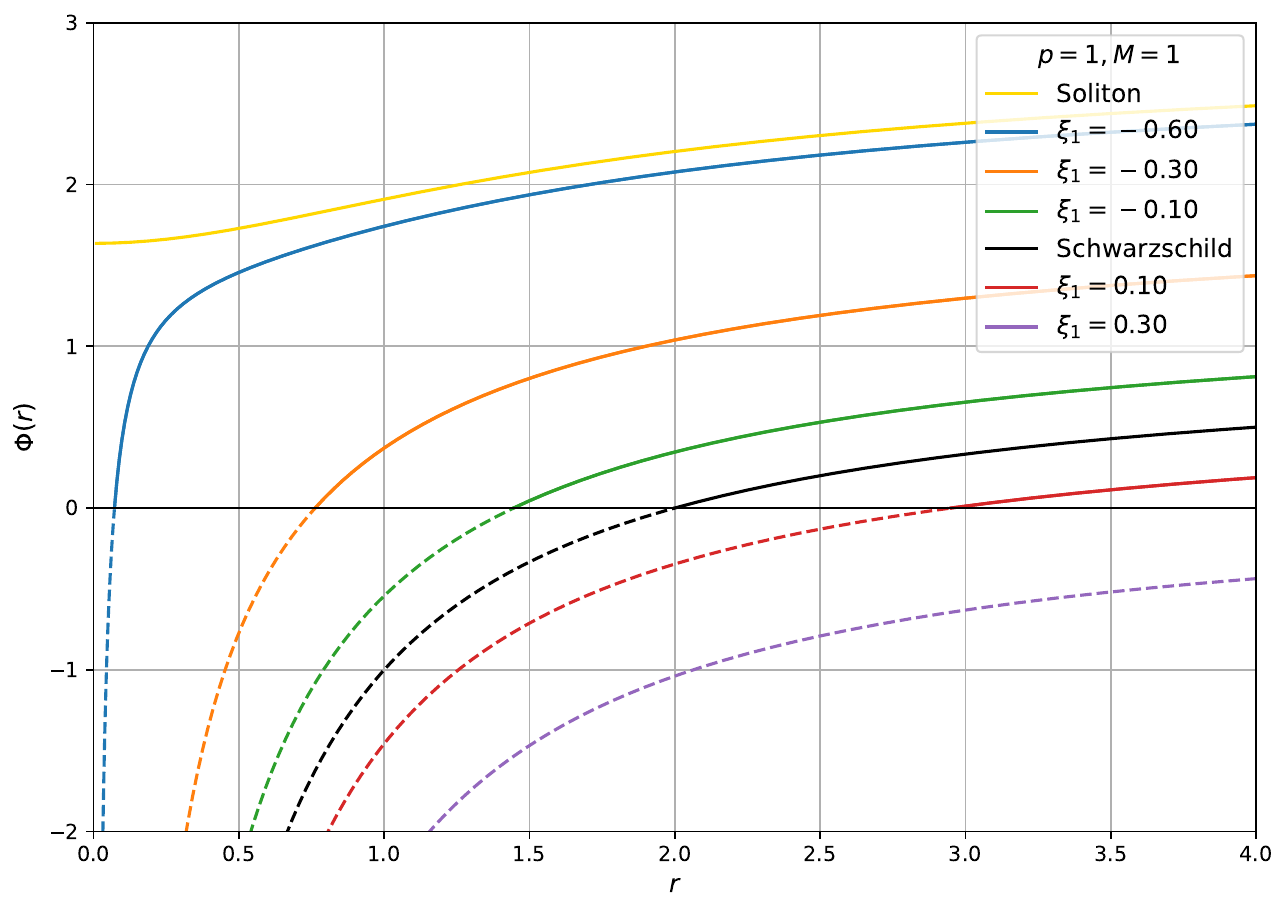}
\includegraphics[width=8.4cm]{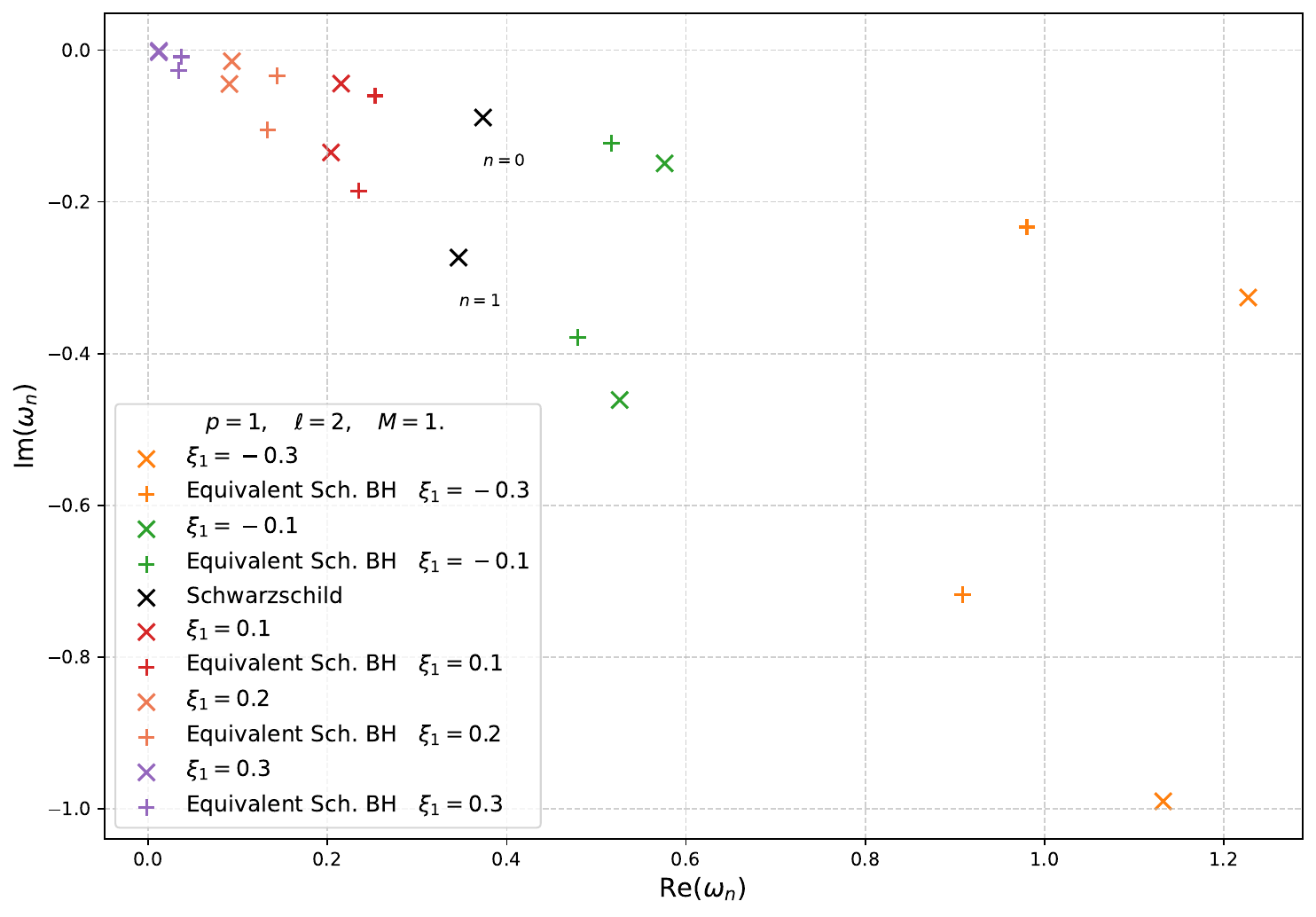}
\end{center}
\caption{Left: Plot of $\Phi(r)$ for $\nth=1$ and $M=1$. We vary the parameter $\xi_{1}$ between $-2M/\pi $ (soliton) and the upper bound $\xi_{1}=1/3$.
The yellow lines correspond to the soliton case while the black lines correspond to the Schwarzschild case.
Right: Plot of the corresponding axial QNM frequencies of the hairy black hole computed via the WKB method (with $\ell=2$). The frequencies computed for the hairy BH are represented by the $\times$ markers while the frequencies computed for a Schwarzschild BH in GR with an equivalent mass are represented by $+$ markers.}
\label{fig:Graphn1}
\end{figure}

\begin{figure}[h]
\begin{center}
\includegraphics[width=8.4cm]{phi_M=2_l=2.pdf}
\includegraphics[width=8.4cm]{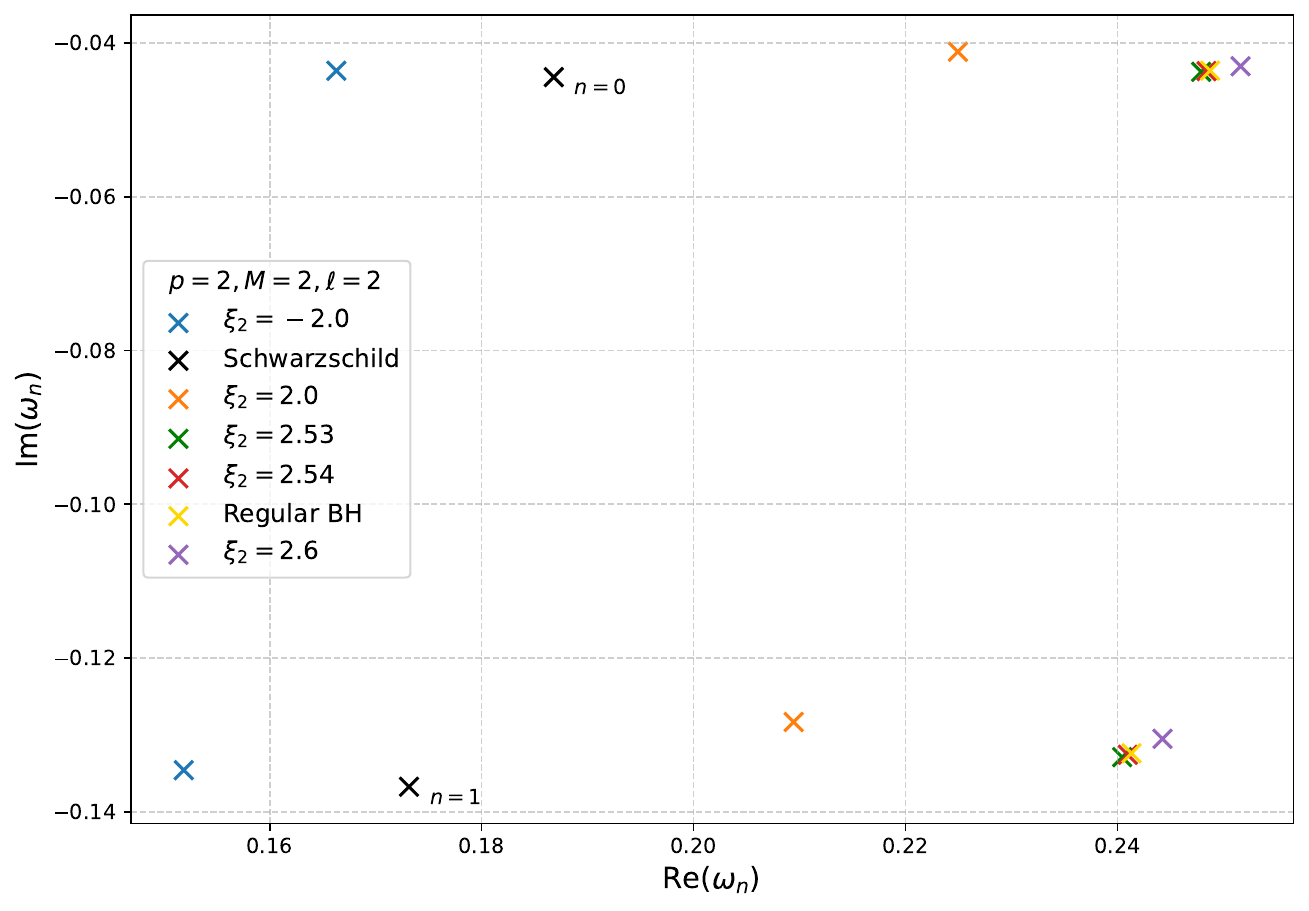}
\end{center}
\caption{Left: Plot of $\Phi(r)$ in the theory $\nth=2$, for $M=2$ and various values of $\xi_2$ between $-2$ and the upper bound $\xi_{2}=3$. The black line corresponds to the GR Schwarzschild BH, and the yellow lines to the special case   of a regular BH.
Right: Axial QNM for the same values of the parameters.}
\label{fig:Graphn2M23}
\end{figure}

\begin{figure}[h]
\begin{center}
\includegraphics[width=8.4cm]{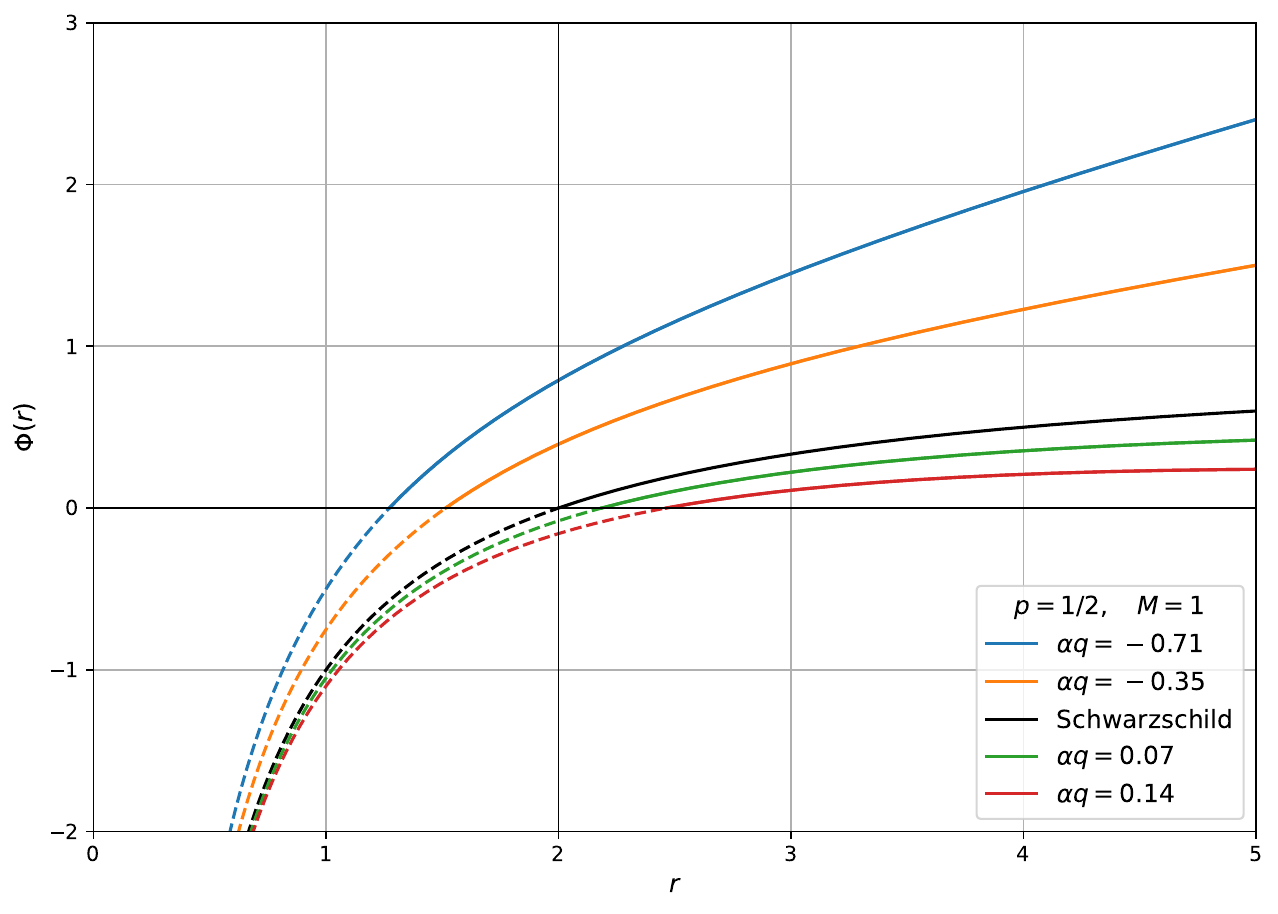}
\includegraphics[width=8.4cm]{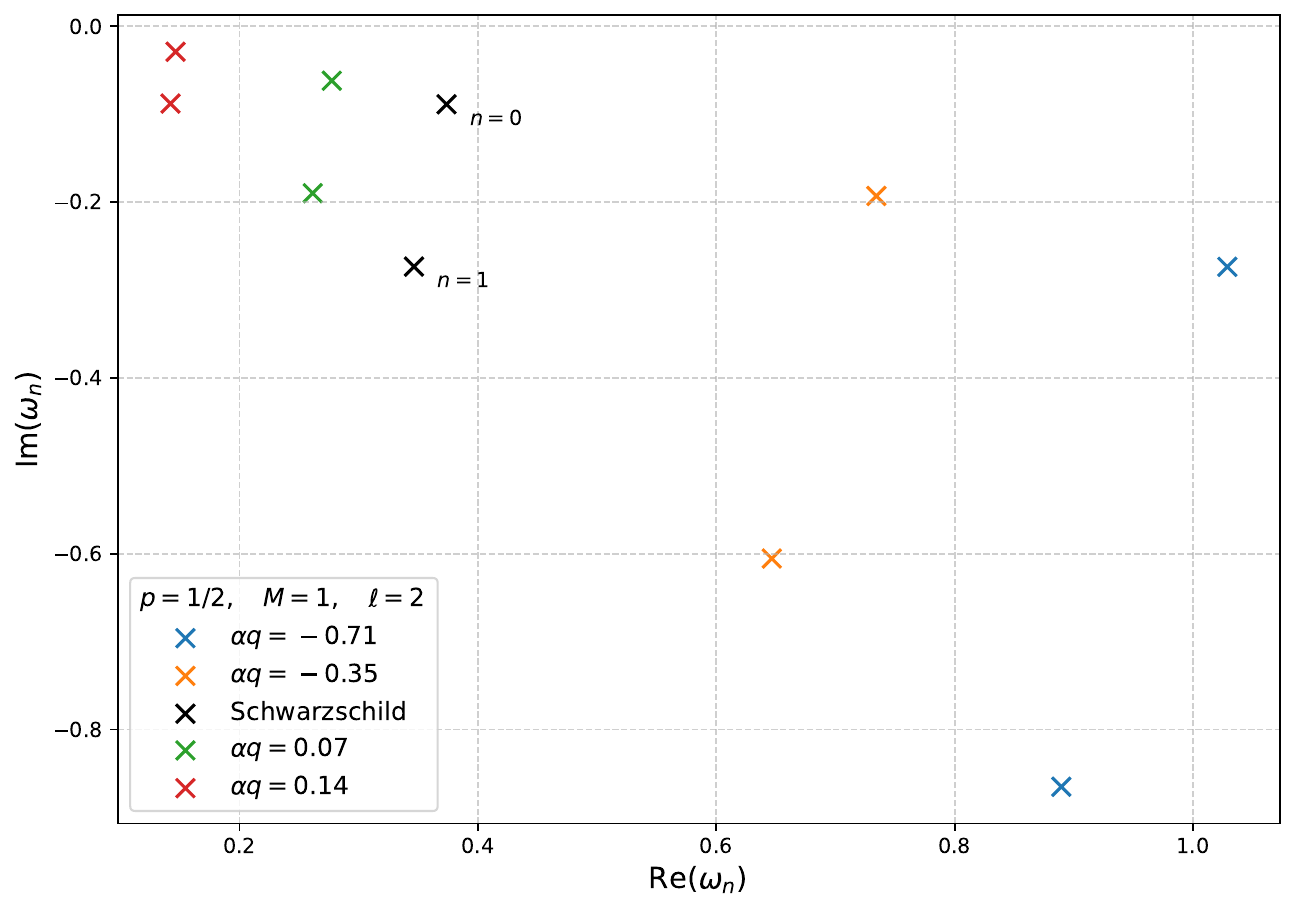}
\end{center}
\caption{Left: Plot of $\Phi(r)$ in the theory $\nth=1/2$, for $M=1$ and various values of $\alpha q$. The black line corresponds to the GR Schwarzschild BH.
Right: Axial QNMs for the same values of the parameters.}
\label{GraphStealth}
\end{figure}

We present our results for the fundamental mode ($n=0$) and the first overtone ($n=1$) in effective potentials with  $\ell=2$,  related to   black hole solutions in the  theories $p=1$ (see Fig.~\ref{fig:Graphn1}), $p=2$ (see Fig.~\ref{fig:Graphn2M23}) and $p=1/2$ (see Fig.~\ref{GraphStealth}).
In these cases, the optimal truncation order is typically  around 5 or 6, similar to what is observed for the Schwarzschild BH in GR, which corresponds to a precision of order $10^{-3}$ for the QNM frequencies.
We have also computed the modes for other values of $\nth$ and  observed  a similar qualitative behavior for the displacement of the QNM frequencies, away from the GR Schwarzschild frequencies,  when varying the deformation parameter $\xi_p$. 

Fig.~\ref{fig:Graphn1} illustrates  the case $ \nth=1 $, for small values of $ \xi_1 $. One can observe that, as $ \xi_1 $ increases, the horizon size $r_g$ (the largest solution of $ \Phi(r) = 0 $) decreases, while both real and imaginary parts of the QNM frequencies increase. This behavior is analogous to that in GR, where the real and imaginary parts of the QNM frequencies scale inversely with the Schwarzschild BH mass, or equivalently, with the horizon radius. However, as shown in Fig.~\ref{fig:Graphn1}, the modes are distinct from those of a Schwarzschild BH in GR with an equivalent mass set to $ r_g/2 $. This is to be expected since the effective metric differs from the GR one.

\medskip

\begin{figure}[h]
\begin{center}
\includegraphics[width=8.4cm]{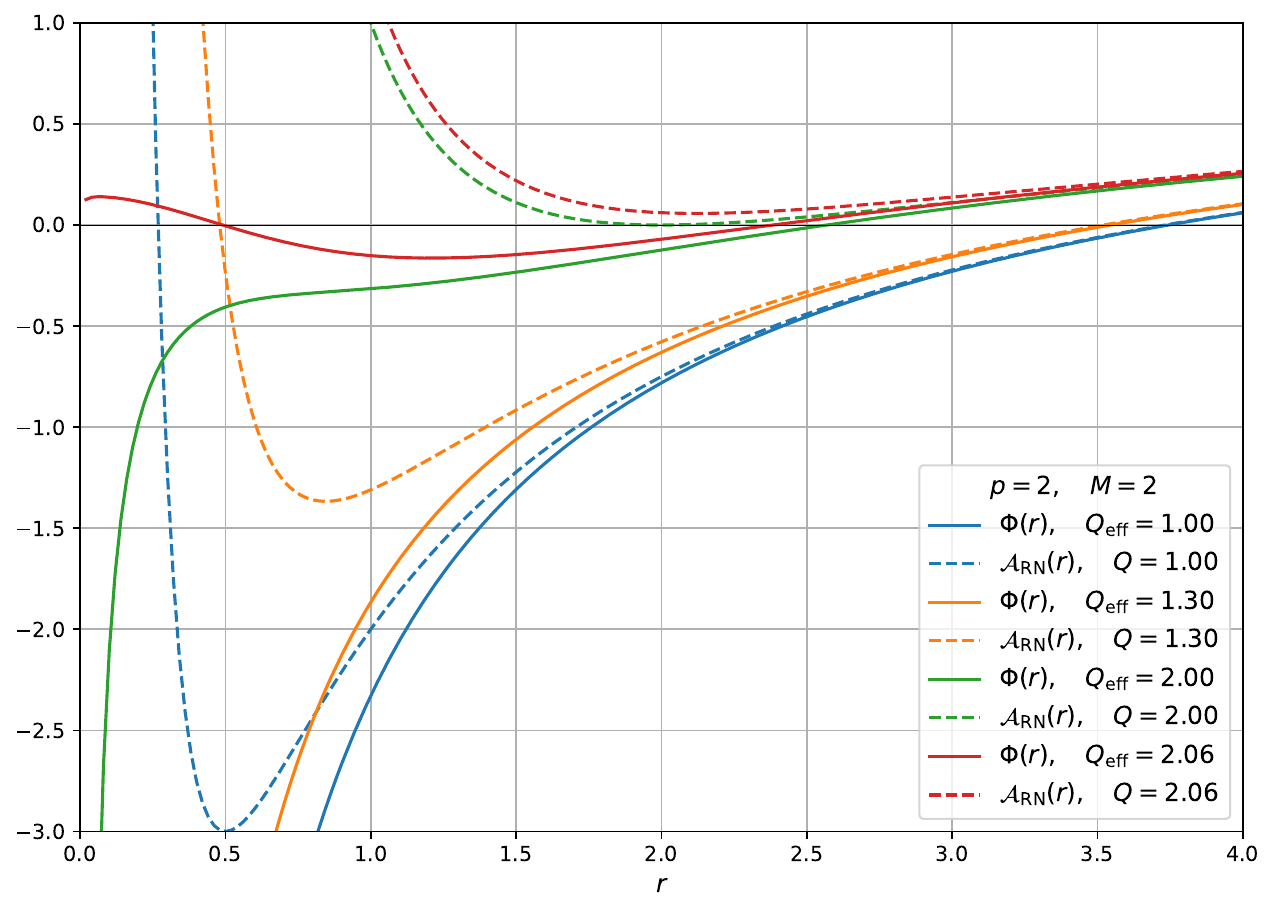}
\includegraphics[width=8.4cm]{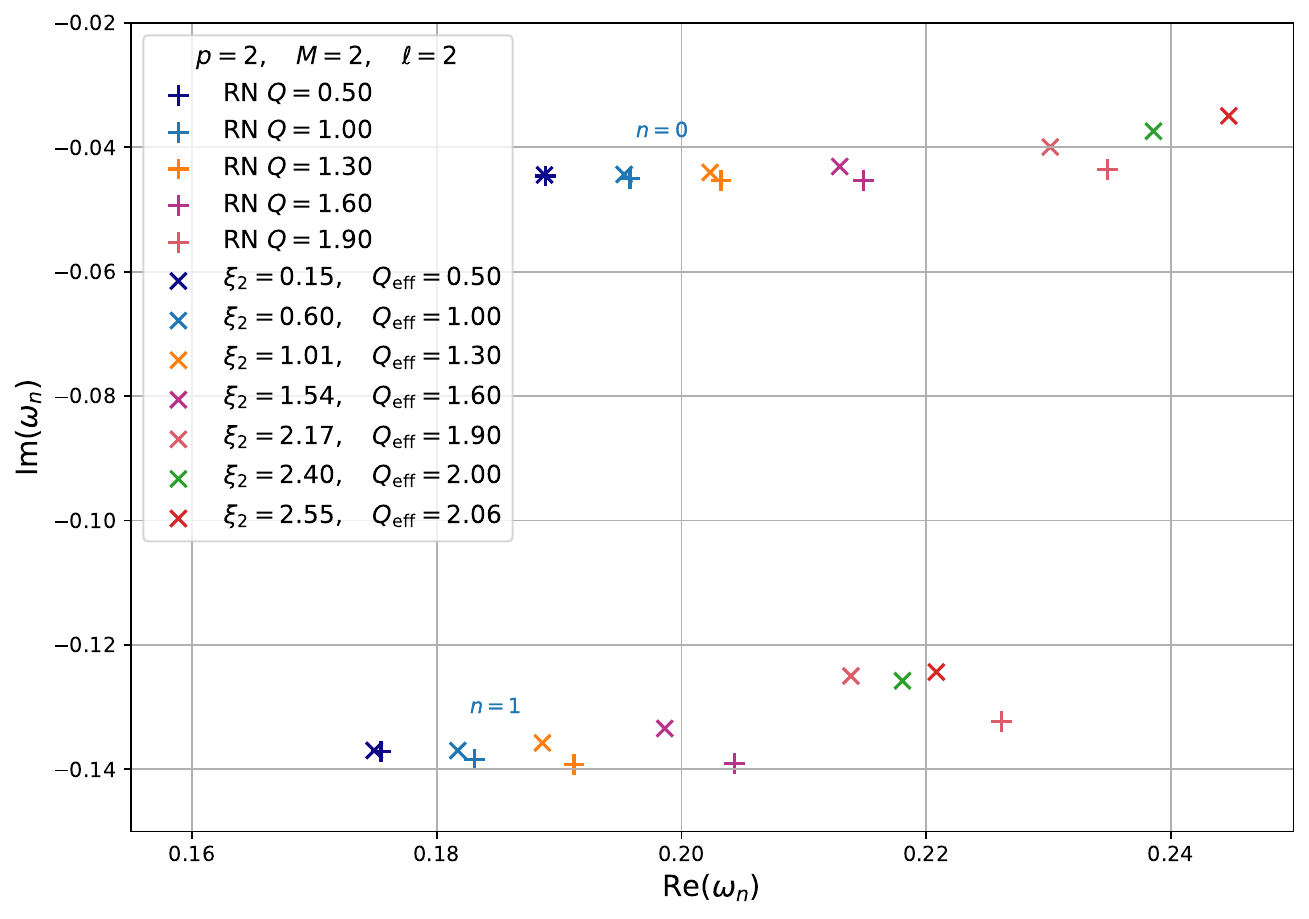}
\end{center}
\caption{Left: Plot of $\Phi(r)$ ($\cA_{\rm RN}(r)  $) for $\nth=2$ and  $M=2$, for various values of the (effective) electric charge $Q$ ($Q_{\rm eff}$).
Right: Axial QNM frequencies of the black hole, for $\nth=2$ and $M=2$. As the electric charge increases the QNMs of the hairy BH and the RN BH are more and more distinct.}
\label{fig:GraphRN}
\end{figure}

Let us now turn to the case $ \nth=2 $,  for relatively large values of $ \xi_2 $, illustrated in Fig.~\ref{fig:Graphn2M23}. 
Interestingly, at large radius, the effective metric closely resembles that of a Reissner-Nordström (RN) black hole. By examining the asymptotic behavior of the metric for $r\gg 1$,  we find 
\begin{eqnarray}
    \mathrm{d} \hat{s}^2 \;= \; -(1-\frac{2M}{r} + \frac{5 \xi_2}{3 r^2}) \, \mathrm{d} t_*^2 + (1-\frac{2M}{r} + \frac{5 \xi_2}{3 r^2})^{-1} \,\mathrm{d} r^2 + r^2 \mathrm{d} \Omega^2 \, + o(1/r^2),
\end{eqnarray}
which enables us to deduce the effective charge via the identification $Q^2_{\rm eff} \equiv 5\xi_2/3$. Comparing  the  QNM  spectrum with that of the corresponding RN black hole,  for different values of $\xi_2$, we find that the first two modes  deviate from their RN counterpart, as shown in Fig.~\ref{fig:GraphRN}. This is expected since the lowest QNMs are primarily influenced by the metric near the peak of the potential, whereas the resemblance to the RN metric holds at large $r$.

Finally, we have also computed the QNM frequencies for the stealth solution, corresponding to the theory $p=1/2$, which were also studied in \cite{Sirera:2024ghv}. In this case, the background solution is Schwarzschild but the effective metric depends on the charge $q$ so that the QNM frequencies migrate in the complex plane when the charge $q$  varies, as shown in Fig.~\ref{GraphStealth}.

\section{Discussion}
In this work, we have considered static black hole solutions in a family of shift-symmetric quadratic DHOST theories \eqref{P_n}. These black holes possess a primary hair,  characterised by an arbitrary parameter independent  of the  mass and   related  to the scalar charge (at least for $\nth>3/2$) arising from the shift symmetry of the theory.
For a  special value of the scalar charge, depending on the mass, given in \eqref{M_reg}, the background solution corresponds to a regular black hole (or a soliton if the metric has no horizon), meaning there is no spacetime singularity at the origin $r=0$. 

We have constructed novel solutions obtained from disformal transformations of these hairy black holes. The disformal transformations we considered fall into two categories: those mapping to a theory where (axial) gravitons propagate at the speed of light ($c_g=c$) and those mapping to a Horndeski theory. The causal structure of the disformed solutions differs from that of the original seed solution. However, they generally (though not always) describe black holes with regularity properties similar to those of the seed black holes, particularly satisfying the same regularity condition \eqref{M_reg}. The comparison between the geometries of  the seed and disformed solutions, as the scalar charge and black hole mass vary, can be nicely illustrated in a phase diagram (see Fig. \ref{fig:Phase_diagram} for theories with $p=2$, when going to the $c_g=c$ frame).

Next, we analysed the dynamics of linear axial perturbations about hairy black holes. Following \cite{Langlois:2022ulw}, we highlighted
the correspondence between their equations of motion and  the dynamics (in the GR sense) of axial perturbations propagating  in an effective metric, which differs from the background metric by a disformal transformation. We then derived, following the same procedure as in GR,  a Schr\"odinger-like  equation for the axial modes in the effective metric, expressed in terms of a tortoise coordinate and  a potential associated with the effective metric. When the potential is positive\footnote{This condition is not necessary as it has been shown in \cite{Kodama:2003kk} (see also \cite{Noui:2023ksf} for applications in DHOST theories).}, this implies that 
the perturbations about the effective metric are stable,
following the argument of Wald in \cite{Wald:1980jn}. 
We then  computed numerically the lowest quasi-normal modes, using the well-known WKB method, for various cases.  We thus could observe how, and in which direction,  the QNMs smoothly deviate from Schwarzschild in the complex plane,  as we vary the mass and scalar charge of the background solution.  Let us also stress that, given a homogeneous BH solutions presented in section \ref{BH_primary_hair}, {\it any} background solution that is related to it by a conformal-disformal transformation, will lead to the same effective metric.

Our analysis of the axial perturbations provides the QNMs associated with the effective metric in the region exterior to the gravitational horizon and one should be careful concerning the implications of our results for the stability of the original background metric. Indeed, the time coordinate $t_*$ that appears in the effective metric differs from the background time coordinate $t$, and one can even notice that the $t_*=$ const hypersurfaces are not spacelike everywhere with respect to the physical metric $g_{\mu\nu}$. In fact, when $q\neq 0$, they always become time-like when one approaches $r_g$. Because of this, a well-posed initial-value problem in the physical metric could become ill-posed in the effective formulation. One must ensure the existence of a common Cauchy hypersurface for both the background and the effective metric. 
Note that this problem is absent when the DHOST theory considered satisfies $A_1=0$, in which  case the background and effective metrics coincide, and therefore $t_*$ and $t$ too. This corresponds to theories where $c_g=c$, for which we have obtained (inhomogenous) black hole solutions.
Apart  from this case, it is thus not fully clear whether the QNMs computed for the effective metric can adequately capture the physical properties of the QNMs associated with the physical black hole. This is an intriguing  question which we plan to address in the future.

More generally, one can wonder which properties of the effective metric are required to guarantee the stability of the hairy BH. 
For example, an open question  is whether a BH solution leading to an effective metric with  a naked singularity  (corresponding to the dotted white region in Fig.~\ref{fig:Phase_diagram}) is pathological or not (see e.g. \cite{Wald:1980jn}). Note that, for black holes obtained in the regularised Einstein-Gauss-Bonnet theory in 4 dimensions  (see the review article \cite{Fernandes:2022zrq} and references within), the effective metric is also that of a naked singularity, as found   in \cite{Langlois:2022eta}, although this time for {\it all} BH solutions, and a similar question arises. 

So far, we have discussed only linear perturbations, which means that photons and gravitons, effectively propagating in different geometries, remain decoupled.  
If these modes were coupled, a superradiance phenomenon could arise, as recently illustrated in a toy model \cite{Cardoso:2024qie}. An instability could also emerge in the analysis of polar perturbations, which contain two coupled degrees of freedom that might propagate in different metrics as well. A final remark on axial modes concerns the eikonal limit when the angular momentum $\ell$ is large. In GR, it is well known that the parameters of quasi-normal modes in this limit are directly related to those of the photon sphere (see {\cite{Cardoso:2008bp}} and references therein). However, this is not the case in tensor-scalar theories, where the effective metric differs from the background one. In this context, QNM parameters are instead associated with the photon sphere of the effective metric, as previously noted in the literature~\cite{Konoplya:2017wot}.

In order to complete our analysis of perturbations of the primary hair black hole solutions, one needs to study the polar (or even) modes of these solutions, which we plan to address in a future work. These modes are technically more challenging, because they involve two degrees of freedom, instead of one in GR, due to the presence of the scalar field.  A preliminary approach could consist in considering only the scalar field perturbation in the fixed background, i.e. treating the scalar field as a  test field, and check  whether such a configuration is stable. To do this, however, one would need to work in the Horndeski frame discussed in section \ref{Section_disformal}, in order to avoid a spurious mode arising from the higher-order scalar field equation of motion in the original frame.

An interesting consequence of the fact that light and gravitational waves propagate in different metrics is the existence of a time delay between gravitational and electromagnetic signals.
Indeed, let us consider an event, of coordinates $t=t_1$ and $r=r_1$, located both outside the luminous and gravitational horizons, at the origin of radial GW  and electromagnetic signals towards a static observer at $r=r_2$. The electromagnetic signal propagates radially in the background metric so that 
    \begin{align}
        t_2^{\rm \ell}-t_1=\int_{r_1}^{r_2}\frac{\mathrm{d} r }{\cA(r)}\,,
    \end{align}
 while the GW signal propagates along a null geodesic in the effective metric (assuming the geometric optics limit applies). Along a null geodesic in \eqref{effectivemetric}, we have $dt_*=(\sqrt{\Ftwo}/\Phi)\mathrm{d} r $ so that, using \eqref{t_star_eff}, we find
    \begin{align}
        t_2^{\rm g}-t_1=\int_{r_1}^{r_2} \, \mathrm{d} r  \, \frac{\sqrt{\Ftwo}+q\psi' A_1}{\Phi}\,.
    \end{align}
Replacing $\Ftwo$  by $1-q^2A_1/(1+r^2)$, according to \eqref{P_n} and \eqref{scalar}, and $A_1$ by $(\cA-\Phi)/q^2$, according to \eqref{coeffs_here}, 
we find that the time delay between the two signals is given by
\begin{align}
    \label{time_delay2}
        t_2^{\rm g}-t_2^{\rm\ell}=\int_{r_1}^{r_2}\mathrm{d} r \left[\left(\frac{1}{\Phi}-\frac{1}{\cA}\right) \sqrt{1-\frac{\cA}{1+r^2}}+ \frac{1}{\Phi}\sqrt{1+\frac{\Phi-\cA}{1+r^2}}-\frac{1}{\cA}\right]\,,
\end{align}
which corresponds, for the observer at $r=r_2$, to the time delay $\Delta\tau_{\rm obs}=\sqrt{\cA(r_2)}(t_2^{\rm g}-t_2^{\rm\ell})$.

To give a simple quantitative estimate of this effect, let us assume
 a small deviation of the background and effective metrics with respect to Schwarzschild, i.e. a small $\xi_p$.  We obtain, at first order in $\xi_p$, the time delay
\begin{align}
        t_2^{\rm g}-t_2^{\rm\ell}\simeq\frac{\xi_p}{2p-1}\int_{r_1}^{r_2} \frac{\mathrm{d} r }{(1+r^2)^{p-1}\cA_s^2}  \left(
        1+\sqrt{1-\frac{\cA_s}{1+r^2}}-\frac{\cA_s}{2(1+r^2)}\right)\,,
    \end{align}
where $\cA_s\equiv 1- 2M/r$ is the metric component for Schwarzschild. Restoring the length scale $\lambda$ and assuming for simplicity $\nth=2$ and $M\gg \lambda$, one gets an expresssion that can be integrated explicitly to yield
\begin{equation}
    t_2^{\rm g}-t_2^{\rm\ell}\simeq \frac23\lambda^2\, \xi_2\left(\frac{1}{r_1-2M}-\frac{1}{r_2-2M}\right)\,.
\end{equation}
If the observer, located at $r_2$, is sufficiently far from the black hole, so that the last term can be neglected, we find
\begin{equation}
    \Delta\tau_{\rm obs}\simeq 1.7\times10^{-6}\, \xi_2\left(\frac{M}{\lambda}\right)^{-2}\left(\frac{r_1}{2M}-1\right)^{-1}\left(\frac{M}{M_\odot}\right)\, {\rm s}\,.
\end{equation}
We see that the effect is very small in general but can be enhanced if the signals originate from a point very close to the outer horizon.  Let us stress that the above formula is valid only if $M\gg \lambda$. Note that if we extrapolate this modified gravity effect to the merger of two neutron stars, this effect is well below the time delay, of the order of the second, measured for the GW170817 event\footnote{Note that, in our case, the modified gravity effect is localised in the vicinity of the compact objects, so there is no cumulative effect on cosmological distances, in contrast with models of modified gravity that try to account for dark energy.}.

\section*{Acknowledgments}
This work was partially supported by the French National Research Agency (ANR) via Grant No. ANR-22-CE31-0015-01 associated with the project Strong. We would like to warmly thank Hugo Roussille for his advice on calculating the QNMs. CC is also very happy to thank George Pappas for interesting discussions on the eikonal limit. DL would like to thank the theory department at CERN for their kind hospitality during the final stage of this work.

\begin{appendices}

\section{Construction of the phase diagram}
\label{App:phase_diagram}
In this Appendix, we give details for the calculation of the phase diagram pictured in Fig. \ref{fig:Phase_diagram}. We are considering only the case $\nth=2$ as the other cases can be done in a similar way.

As we have shown in the paper, most of the geometrical properties of background and effective metrics can be captured by the respective functions,
\begin{eqnarray}
    {\cA}(r) & = &1 - \frac{2M}{r} + \xi_2 \left( \frac{\pi/2 - \arctan(r)}{r} + \frac{1}{1+r^2} \right) \, , \\
    {\Phi}(r) & = & \cA(r) - \frac{\xi_2}{3} \frac{1}{1+r^2} \, .
\end{eqnarray}

\subsection{Regular solutions}
First, we recall that the metric is regular at the origin $(r=0)$ when 
\begin{eqnarray}
    \xi_2 \; = \; \frac{4}{\pi} \, M_2^{\rm reg} \, ,
\end{eqnarray}
which corresponds to a line in the diagram. Along this line, we could encounter two different possibilities for the background and the effective metric, either they correspond to a regular black hole or to a soliton. We are at the limit between these two solutions when $\xi_2$ (or equivalently $M$) allows the following system of equations to admit a unique solution for $r$,
\begin{eqnarray}
    &&\cA(r)= \cA'(r)=0 \, \qquad (\text{for the bakground metric}), \\
    &&\Phi(r)=\Phi'(r)=0 \,  \qquad (\text{for the effective metric}).
\end{eqnarray}
The equation $\cA'(r)=0$  (or $\Phi'(r)=0 $) shows that the solution for $r$ does not depend on $\xi_2$ and its value can be computed numerically, and we obtain
\begin{eqnarray}
    r \simeq 1.825 \quad \text{(background)} \, , \qquad   r \simeq 1.210 \quad \text{(effective metric)} \, .
\end{eqnarray}
We substitute these values to the remaining equations $\cA(r)=0$ or $\Phi(r)=0$ and we obtain the value of $\xi_2$ which gives the limit between the two regular solutions: $\xi_{2,{\rm sol}} \simeq 2.816$ (or equivalently $M^{\rm reg,II}_2 \simeq 2.212$) for the background metric while $\xi_{2,{\rm sol}} \simeq 2.189$ (or equivalently $M^{\rm reg,I}_2 \simeq 1.719 $) for the effective metric. 
Hence, when $\xi_2$ is below this value, the solution is a soliton while it is a regular black hole when it is above.

\medskip

A similar analysis can be carried out for other values of $ n $. In particular, when $p = {5}/{2} $, an exact value for $ \xi_{5/2,{\rm sol}} $ can be determined. Specifically, one observes that the derivative $ \mathcal{A}'(r) $ in equation \eqref{A_n52} vanishes at $ r = \sqrt{2} $, and subsequently, it can be shown that $ \mathcal{A}(\sqrt{2}) = 0 $ when $ \xi_{5/2,{\rm sol}} = {9 \sqrt{3}}/{4} $.

\subsection{Non regular solutions}
Now we study the case where the solution is not regular at the origin. We first focus on the background metric.  The study of the effective metric is very similar and will be shortly discussed.

In the case where $\xi_2 <4M/\pi$ (below the line we have studied in the previous subsection), the solution is always a black hole. However, it could have one or three horizons.   As in the previous subsection, we are at the limit between these two possibilities when the following system of equations,
\begin{eqnarray}
\label{AAprime}
    \cA(r) \;  = \; \cA'(r) \; = \; 0 \, ,
\end{eqnarray}
admits one (unique) solution for $r$. If such a solution exists, then we get a relation between $M$ and $\xi_2$. 

As $r$ is necessarily different from zero, the equations \eqref{AAprime} imply 
\begin{eqnarray}
    \frac{\dd (r \cA )}{\dd r} \; = 1 + \xi_2 \frac{\dd }{\dd r} \left( \frac{\pi/2 - \arctan(r)}{r} + \frac{1}{1+r^2} \right)\; = \;0  \, .
\end{eqnarray}
This leads to
\begin{eqnarray}
\label{r4r2}
    r^4 +2(1-\xi_2) r^2 + 1 \, = \, 0 \, ,
\end{eqnarray}
which admits solutions only if $\xi_2 \geq 2$. When $\xi_2 >2$, it has the following two distinct solutions,
\begin{eqnarray}
    r_\pm(\xi_2) \; = \; \sqrt{\xi_2-1 \pm \sqrt{\xi_2(\xi_2-2)}} \, , 
\end{eqnarray}
which merge into the unique solution $r_\pm=1$ when $\xi_2=2$.

Then, one substitutes these solutions into the equation $\cA(r)=0$ and we obtain the two branches, that can be seen in the
phase diagram, defined by the following relation between $M$ and $\xi_2$,
\begin{eqnarray}
    2M \; = \;  \left[ 1+ \xi_2 \left( \frac{\pi/2 - \arctan(r_\pm)}{r_\pm} + \frac{1}{1+r_\pm^2} \right)\right] r_\pm \,\, .
\end{eqnarray}

As a consequence, between the two branches and below the regularity line (i.e. $\xi_2 <4M/\pi$), the solution is a black hole with three horizons, otherwise it is a black hole with only one horizon. 

When the upper branch crosses the regularity branch, the nature of solution changes. Indeed, in the domain $\xi_2  > 4M/\pi$, there are two zones: above the upper branch, the solution is a naked singularity while it is a black hole with two horizons below this  branch.

We can easily reproduce the  very similar analysis for the effective metric, in which case the frontier between a black hole with one horizon and a black hole with three horizons (in the regime where $\xi_2  < 4M/\pi$) can be obtained by solving
\begin{eqnarray}
    \Phi(r) \; = \; \Phi'(r) \; = \; 0 \, .
\end{eqnarray}
We follow the same strategy as for the background metric and \eqref{r4r2} is now replaced by
\begin{eqnarray}
    3r^4 +(6-5\xi_2) r^2 + 3-\xi_2 \; = \; 0 \, .
\end{eqnarray}
It admits solutions only if $\xi_2 \geq 48/25$, in which case the solutions for $r$ are
\begin{eqnarray}
    r_\pm(\xi_2) \; = \; \frac{1}{\sqrt{6}}\sqrt{5\xi_2-6 \pm \sqrt{\xi_2(25\xi_2-48)}} \, ,
\end{eqnarray}
which merge into a unique solution for $\xi_2 = 48/25$. The rest of the analysis is strictly the  same as the previous ones with those expressions of $r_\pm$.

\section{Disformal transformations to the Horndeski frame}
\label{App:Horndeski}

As discussed in the main text, it is possible to perform a purely disformal transformation \eqref{disformal_transf}, i.e. with $C=1$, of the original metric to map one of the theories \eqref{P_n} into a Horndeski theory. 
In the case $\nth=2$, the required disformal factor is given by
\begin{equation}
    \label{D_H_p2}
    \cDis(X)=\frac{3\alpha}{2} X\,,
\end{equation}
according to \eqref{D_Horndeski}.
Substituting the above expression into the relation \eqref{disformed_X} between $X$ and $\tilde{X}$, one finds that the radial dependence of the kinetic term in the new frame, which we will denote here $\XH$, is given by
\begin{equation}
\XH(r)=\frac{q^2(1+r^2)}{2\left[(1+r^2)^2-\xi_2\right]}\,,
\end{equation}
where we have used the definition \eqref{xi} of $\xi_2$.

Interestingly, in the present case one can invert  the relation \eqref{disformed_X} in order  to express  the  new DHOST functions in terms of $X_{\rm H}$, thus enabling us to define explicitly the corresponding Horndeski theory. In general, the new theory can be defined only implicitly.
Since \eqref{disformed_X} with \eqref{D_H_p2} yields a quadratic equation for $X$, we obtain  two possible solutions for $X$ in terms of $X_{\rm H}$. Choosing the appropriate solution, depending on  the sign of $\alpha$ so that $X$ and $X_{\rm H}$ share the same sign, we find for the two DHOST functions characterising the Horndeski theory: 
\begin{align}
    &\Ftwo^{\rm H}(\XH)=\frac{12\alpha\XH^2-1+\sqrt{1+12\alpha\XH^2}}{3\sqrt{6\alpha}\XH\sqrt{\sqrt{1+12\alpha\XH^2}-1}}\,,\quad
    A_1^{\rm H}(\XH)=-\frac{\left(\sqrt{1+12\alpha\XH^2}-1\right)^{3/2}}{6\sqrt{6\alpha}\XH^2}\quad (\alpha>0)
\end{align}
or
\begin{align}
    \Ftwo^{\rm H}(\XH)&=\frac{1+12|\alpha|\XH^2+\sqrt{1+12\alpha\XH^2}}{3\sqrt{6|\alpha|}\XH\sqrt{\sqrt{1+12\alpha\XH^2}+1}}\,,\qquad
    A_1^{\rm H}(\XH)=\frac{\left(\sqrt{1+12\alpha\XH^2}+1\right)^{3/2}}{6\sqrt{6|\alpha|}\XH^2}\quad (\alpha<0)\,.
\end{align}

\section{Formulas with the convention $\bar X=\partial_\mu\phi\, \partial^\mu\phi$}
\label{App:X_other}
In the present work, we have defined the kinetic term as $X=-\frac12 \partial_\mu\phi \, \partial^\mu\phi$ but  many papers  adopt instead the convention 
\begin{equation}
\label{Xbar}
    \bar X\equiv \partial_\mu\phi\,\partial^\mu\phi\,,
\end{equation}
where we use here a bar to distinguish this definition from the convention used in the main text. In order to make the comparison  or use of our results easier, we present below some of our key formulas in this alternative convention. 

With the convention (\ref{Xbar}), beyond Horndeski theories are characterised by the conditions
\begin{eqnarray}
A_2=-A_1   \,, \qquad A_4=-A_3= \frac{2}{\Xb}\left(2\Ftwo_{,\Xb}-A_1\right) \,, \qquad A_5=0\,,
\end{eqnarray}
while  the Horndeski theories verify
\begin{eqnarray}
A_2=-A_1= -2\Ftwo_{,\Xb}  \,, \qquad A_3=A_4=A_5=0\,.
\end{eqnarray}
The theories we have considered in this work  are described by the functions 
\begin{align}
\label{theories}
    P&=-\frac{\alpha }{2^{\nth-1} \lambda^2} (-\Xb)^\nth\, , \quad
    \Ftwo=1+\Xb A_1(\Xb)\, , \quad
    A_1=\frac{\alpha}{2^\nth} (-\Xb)^{\nth-1}\,, \quad 
A_3=\frac{\alpha}{2^{\nth-1}}(2\nth-1) (-\Xb)^{\nth-2}\,,
\qquad
\end{align}
while $A_2=-A_1$, $A_4=-A_3$ and $A_5=0$. They all satisfy  $Z=-\Ftwo+\Xb A_1=-1$.
For all BH solutions, the radial profile of the kinetic time is given by
\begin{equation}
\Xb=-\frac{q^2}{1+(r / \lambda)^2}\,.
\end{equation}

The disformal transformation of $\Ftwo$ and $A_1$ read
\begin{eqnarray}
\tilde \Ftwo & = & 
\frac{\Ftwo}{\cCo (1 +\Xb \cDis/\cCo)^{1/2}}
 \, ,\\
\tilde A_1 & = & \left( 1 +\Xb {\cDis}/{\cCo} \right)^{3/2} (A_1 - \frac{\cDis}{\cC+ \Xb\cDis } \Ftwo) 
\,,
\end{eqnarray}
while $\tilde{\Xb}$ is given by
\begin{align}
    \tilde{\Xb}=\frac{\Xb}{\cCo+\Xb\cDis}\,,
\end{align}
requiring  the condition $C+\Xb D>0$.

To reach a theory such that $c_g=c$, one needs a disformal transformation such that
\begin{align}
    \frac{\cDis}{\cCo}=\frac{A_1}{\Ftwo-\Xb A_1}\,,
\end{align}
which, for our specific theories \eqref{theories}, reduces to the condition ${\cDis}/{\cCo}=A_1$.
Finally, one  can obtain  a Horndeski theory, satisfying $\tilde A_1=2 \tilde \Ftwo_{\tilde{\Xb}}$, via a  purely disformal transformation with
\begin{align}
    \cDis(\Xb)=\frac{(2\nth-1)\alpha}{2^\nth(\nth-1)}(-\Xb)^{\nth-1}\quad (\text{for} \; \nth\neq 1)\,.
\end{align}

\end{appendices}

\bibliographystyle{utphys}
\bibliography{References}

\end{document}